\documentclass[aps,pra,preprint,showpacs,amsfonts]{revtex4}

\usepackage{epsfig}

\begin{document}

\section*{RPA GREEN'S FUNCTIONS OF THE ANISOTROPIC HEISENBERG MODEL}

\centerline{Andre Johannes Stoffel and Mikl\'{o}s Gul\'{a}csi}

\centerline{\sl{Max-Planck-Institute for the Physics of Complex Systems, 
D-01187 Dresden, Germany}}
\centerline{\sl{Nonlinear Physics Centre, Australian National University,
Canberra, ACT 0200, Australia}}

\centerline{\today}

We solve in random-phase approximation the anisotropic Heisenberg model, 
including nearest and next-nearest neighbour interactions by calculating 
all Green's functions and pair correlation functions in a cumulant 
decoupling scheme. The general exposition is pedagogic in tone and is 
intended to be accessible to any graduate student or physicist who 
is not an expert in the field. 


\section{Introduction}
\label{sec:intro}

Recently we analysed the properties of an anisotropic Heisenberg 
model in an external longitudinal field on a bcc lattice, with a 
particular application to supersolids \cite{epl,epjb1,epjb2}. 
However, detailed derivation of the Green's functions and correlation 
functions have not yet been published, as such we fill 
this gap hereafter. For the interested reader we present in 
Appendix A the connection between the model analysed hereafter 
and ${}^4$He. 

\begin{figure}[t]
\centering
\includegraphics[width=7cm]{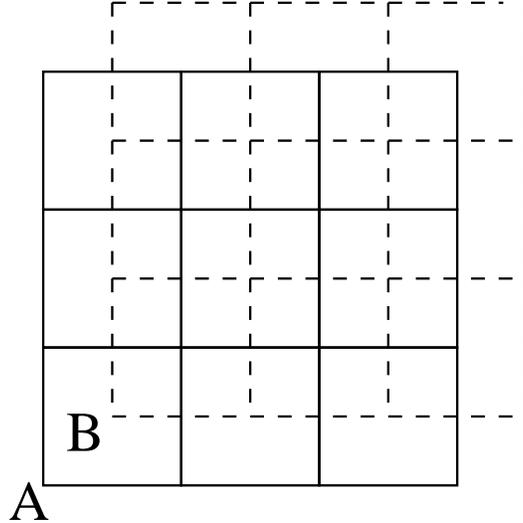} 
\caption{\label{fig:fig1} The bcc lattice consists of two interpenetrating 
sc sub-lattices, i.e., sub-lattice A and sub-lattice B. For simplicity we 
only have drawn the two 
dimensional case.}
\end{figure}

We work with the standard anisotropic Heisenberg model,
defined by the Hamiltonian: 
\begin{equation}
H=h^z \sum_i S^z_i+\sum_{ij}J^{\|}_{ij}S^z_iS^z_j+
\sum_{ij}J^{\top}_{ij}(S^x_i S^x_j+S^y_iS^y_j) 
\label{new_one}
\end{equation}
on a bcc lattice shown in Figure (\ref{fig:fig1}). 
It is known that the Hamiltonian defined in Equation (\ref{new_one}) gives 
four magnetic phases: ferromagnetic, canted ferromagnetic, canted 
anti-ferromagnetic and anti-ferromagnetic phases which we abbreviate 
by FE,CFE, CAF and AF, see Table (\ref{tab:one}). 
The order parameters, $m_1$ for off-diagonal long range order and  $m_2$
for diagonal long range order in the magnetic system are defined by:
\begin{eqnarray}
m_1=\langle S^x_A\rangle+\langle S^x_B\rangle \nonumber \\
m_2=\langle S^z_A\rangle-\langle S^z_B\rangle
\end{eqnarray}
In the following we will use these order parameters to identify 
the phases within the random-phase approximation.
\begin{table}
\centering
\begin{tabular}{|c|c|c|c|}
\hline
&&&\\
Spin Configuration& Magnetic Phase & ~~~ ODLRO ~~~ & ~~~~ DLRO ~~~~ \\
&&&\\
\hline
&&&\\
$\uparrow\uparrow$ & FE & No & No \\
&&&\\
$\nearrow\nearrow$ & CFE & Yes & No \\
&&&\\
$\nearrow\swarrow$ & CAF  & Yes & Yes \\
&&&\\
$\uparrow\downarrow$& AF & No & Yes \\
&&&\\
\hline
\end{tabular}
\caption{\label{tab:one} Possible magnetic phases of the anisotropic Heisenberg
model. All phases are defined by their long range order, i.e., 
off-diagonal long-range order (ODLRO) and diagonal long-range order (DLRO).}
\end{table}

The Heisenberg model has been studied thoroughly  although there exists 
only a classical mean-field solution for the canted ferromagnetic and 
anti-ferromagnetic phases \cite{Fisher,LiuFisher}. Hence, our interest 
is in obtaining a better than mean-field solution of these canted phases. We used 
\cite{epl,epjb1,epjb2} the equation of motion technique for the 
time-temperature-dependent Green's function \cite{AGD}, 
$G^{xy}_{ij}(t)_{Ret/Adv}=\mp i\theta(\pm t)\langle[S^x_i(t),S^y_j]\rangle ]$, 
for all four phases, which we solve in random-phase approximation (RPA).
In order to maintain pair 
correlations as accurate as possible, we have chosen a cumulant decoupling 
scheme to approximate higher order Green's functions beyond the random-phase 
approximation. This is the first time that a thorough cumulant random-phase 
approximation many-body calculation has been solved for the canted phases and 
as such it represents a vital step in the 
understanding the behaviour of the three dimensional anisotropic Heisenberg
model in the presence of an external magnetic field. Details of the Green's functions
are given in the next Chapter. 

By using these Green's functions we calculate exactly all the pair correlations,
also given in the next Chapter, where we show that 
this decoupling scheme couples six Green's functions, one for each spin component 
in $x$, $y$ and $z$ direction on the two sub-lattices respectively, to a set of 
six equations. Due to the enormous number of terms (1024 in total) within these 
Green's functions we have decided to reproduce their exact form only at the end. 
Using these Green's functions 
it can be shown or alternatively argued that the Goldstone theorem of gapless 
modes imposes an additional condition \cite{Nafari} on the mean fields of
the canted phases, reducing the number of order-parameters by two. 
As this condition does not apply to the ferromagnetic and anti-ferromagnetic 
phases, their Green's functions are structurally different.

For completeness we also re-derive the classical mean-field Green's functions 
in Appendix B. While in Appendix C we establish a link between the classical 
mean-field and the random-phase approximations. 

\section{Green's Functions}
\label{sec:GF}

The most versatile and successful method of solving
many-body problems involve the concept of Green's function. 
The Matsubara formalism \cite{AGD} for imaginary time Green's functions
at finite temperature is 
commonly used to analyze fermionic and bosonic
many-body systems. 
However, in spin systems a different type of 
Green's function is primarily applied as this
makes it easier to deal with the algebraic properties
of spin operators.
The retarded and advanced  Tyablikov \cite{Bogo,Tyablikov} commutator Green's 
function at finite temperature defined in real time are:
\begin{eqnarray}\label{GFdef}
G^{\mu\nu}_{ij_{Ret}}(t)=-i\theta(t)\langle [S^{\mu}_i(t),S^{\nu}_j]  \rangle \nonumber \\
G^{\mu\nu}_{ij_{Adv}}(t)=i\theta(-t)\langle [S^{\mu}_i(t),S^{\nu}_j]  \rangle
\end{eqnarray}
The average $\langle \rangle$ involves the usual quantum mechanical
as well as thermal averages and
$\mu$ and $\nu$ are elements of $\{x,y,z\}$ and $i$ and $j$
denote the lattices sites.
There is no Wick's theorem for 
spin systems and therefore no perturbative approaches by 
means of Feynman diagrams are available. However this
is not necessarily a disadvantage as the method of
the equation of motion is equally powerful in many cases
and often much more compact.
The basic idea of this method  
is to find a linear differential equation for the Green's function.
Therefore we can differentiate the $xy$-Green's function with respect
to time:
\begin{eqnarray}\label{eq_of_m}
i \partial_t G^{xy}_{ij_{Ret}}(t)= \delta(t)\langle[S^x_i,S^y_j]\rangle
-i\theta(t)\langle[[S^x_i,H],S^y_j]\rangle \nonumber \\
i \partial_t G^{xy}_{ij_{Adv}}(t)= \delta(t)\langle[S^x_i,S^y_j]\rangle
+i\theta(-t)\langle[[S^x_i,H],S^y_j]\rangle \nonumber \\
\end{eqnarray}
Here we have used the Heisenberg equation of motion $i  \partial_t S^x=[S^x_i,H]$
for the operator $S^x$.
For the anisotropic Heisenberg model the commutator yields:
\begin{eqnarray}
[S^x_i,H]=-i\mu S^y_i-2i\sum_jJ^{\|}_{ij}S^y_i S^z_j+
                      2i\sum_j J^{\top}_{ij}S^z_i S^y_j
\end{eqnarray}
Consequently the right hand side of Equation (\ref{eq_of_m}) involves
higher, third order Green's functions. In principle those third
order functions can be expressed in terms of even higher order.
This procedure will go on indefinitely yielding 
a series of infinite order. The central idea is 
to introduce a suitable approximation which truncates 
this series. 
Here, in order to keep the complexity to a manageable level we 
 split up the three operator correlation functions. 

The cumulant decoupling \cite{Brown} is based on the assumption that
the last term of the following
equality is negligible:
\begin{eqnarray}
\lefteqn{\langle \hat{A}\hat{B}\hat{C} \rangle =} \nonumber\\
&&\langle \hat{A}\rangle\langle \hat{B}\hat{C} \rangle +
\langle \hat{B}\rangle\langle \hat{A}\hat{C} \rangle \nonumber\\ 
&&+\langle \hat{C}\rangle\langle \hat{A}\hat{B} \rangle 
-2\langle \hat{A}\rangle\langle \hat{B}\rangle\langle\hat{C} \rangle
\nonumber \\
&&+\langle (\hat{A}-\langle \hat{A} \rangle) 
(\hat{B}-\langle \hat{B} \rangle)
(\hat{C}-\langle \hat{C} \rangle) \rangle
\end{eqnarray}
This approximation is justified if the quantum fluctuations 
are small and do not deviate far from their mean field values,
which is the case in three dimensions \cite{Auer}.
As a consequence the third order correlation functions
split into product terms of a second order correlation functions and
the mean-field of a single operators:
\begin{eqnarray}
\lefteqn{\langle[[S^{\alpha}_i(t),S^{\beta}_k(t)],S^y_j]\rangle
 \rightarrow} \nonumber\\
&&\langle S^{\alpha}_i(t)\rangle \langle [S^{\beta}_k(t),S^y_j]\rangle
+\langle S^{\beta}_k(t)\rangle \langle [S^{\alpha}_i(t),S^y_j]\rangle
\end{eqnarray}
Finally we obtain a differential equation which only involves 
second order Green's functions.
\begin{eqnarray}
\lefteqn{i \partial_t G^{xy}_{ij}(t)=}\nonumber\\
&&i  \delta(t) \delta_{ij} \langle S^z_i \rangle
-i\mu G^{yy}_{ij}(t)\nonumber\\
&&-2i\sum_lJ^{\|}_{il}
           (\langle S^y_i(t)\rangle G^{zy}_{lj}(t)+
            \langle S^z_l(t)\rangle G^{yy}_{ij}(t)) \nonumber \\
           &&+2i\sum_lJ^{\top}_{il}
           (\langle S^z_i(t)\rangle G^{yy}_{lj}(t)+
            \langle S^y_l(t)\rangle G^{zy}_{ij}(t)) 
\end{eqnarray}
Here we have dropped the subscripts for the retarded and the advances
Green's functions since the equations equally hold 
for both. In the next Chapter we will see that 
the  advanced and the retarded Green's functions
are actually represented by the same Fourier transform; 
the difference comes from the path along which the
inverse Fourier integral is carried out.
Similar relations also exist for the time derivatives of 
$ G^{yy}_{ij}(t)$ and $ G^{zy}_{ij}(t)$. Together these
equations form a closed set of linear differential equations.
Usually linear differential equations are readily solved 
by Fourier Transformation. The time-frequency Fourier and inverse
Fourier transforms 
are given by:
\begin{eqnarray}
G^{xy}_{ij}(t)=\int d\omega e^{-i\omega t}
G^{xy}_{ij}(\omega) \\
G^{xy}_{ij}(\omega)=\frac{1}{2\pi}\int dt e^{i\omega t}
G^{xy}_{ij}(t) 
\end{eqnarray} 
One might think that due to broken translational symmetry of the 
canted anti-ferromagnetic and the anti-ferromagnetic phases, featuring 
diagonal long-range order, a spacial Fourier
transform might not be applicable. Actually, the canted anti-ferromagnetic 
and anti-ferromagnetic states 
do exhibit discrete translational symmetry, namely they are invariant under
$r_{ij}\rightarrow r_{ij}+n_1 a_1+n_2 a_2+n_3a_3$, where 
$a_1$, $a_2$ and $a_3$ are the basic lattice vectors of the sc sub-lattice and
 $n_1,n_2,n_3\in Z$ . This translation maps each sub-lattice onto itself and
the number of equation doubles as we have to treat each sub-lattice
separately. The Fourier transform into k-space is defined by:
\begin{eqnarray}
G^{xy}_{ij}(\omega)=\int d^3k  e^{i k R_{ij}}
G^{xy}(k,\omega)\nonumber\\
G^{xy}(k,\omega)=\frac{1}{(2\pi)^3 n}\sum_{j}  e^{-i k R_{ij}}
G^{xy}_{ij}(\omega) 
\end{eqnarray}
After successively carrying out time and space Fourier transforms
we derive a set of six algebraic equations, determining six Green's functions
that represent the xy-,yy-, and zy-spin correlations on 
each sub-lattice. The detailed calculation is carried out at the end of this Chapter.
In matrix form this set of equations reads: 
 \begin{equation}\label{Meq}
 M \cdot \gamma= v
 \end{equation}
where 
\begin{eqnarray}
M=\left(
\begin{array}{rrrrrr}
 i\omega & 0    &M_{13}&M_{14}&M_{15}&M_{16} \\
 0     & i\omega&M_{23}&M_{24}&M_{25}&M_{26} \\ 
-M_{13}&-M_{14}&i\omega &0&M_{35}&M_{36} \\
-M_{23}&-M_{24}&0&i\omega &M_{45}&M_{46} \\
 M_{51}& M_{52}&M_{53}&M_{54}&i\omega &0 \\
 M_{61}& M_{62}&M_{63}&M_{64}&0&i\omega \\
\end{array}
\right)
\end{eqnarray}
and
\begin{equation}
\gamma= \left(\begin{array}{llllll}  
{G_a}_k^{xy}(\omega)\\
{G_b}_k^{xy}(\omega)\\ 
{G_a}_k^{yy}(\omega)\\
{G_b}_k^{yy}(\omega)\\
{G_a}_k^{zy}(\omega)\\
{G_b}_k^{zy}(\omega)
 \end{array}\right)
\end{equation}
as well as
\begin{equation}
v=\frac{1}{(2\pi)^4} \left(\begin{array}{llllll}  
\langle S^z_A\rangle\\
\langle S^z_B\rangle\\ 
0\\
0\\
\langle -S^x_A\rangle\\
\langle -S^x_B\rangle
 \end{array}\right)
\end{equation}
The components of the matrix M are given by:
\begin{eqnarray}
&&M_{13}= 2h^z+4\langle S^z_A\rangle (J_2^{\|}(0)-J_2^{\top}(k))+4\langle 
        S^z_B\rangle J_1^{\|}(0)\nonumber\\ 
&&M_{14}=-4\langle S^z_A\rangle J_1^{\top}(k) \nonumber\\
&&M_{23}=-4\langle S^z_B\rangle J_1^{\top}(k) \nonumber\\
&&M_{24}=2h^z+4\langle S^z_B\rangle (J_2^{\|}(0)-J_2^{\top}(k))+4\langle 
        S^z_A\rangle J_1^{\|}(0)\nonumber\\
&&M_{15}=4\langle S^y_A\rangle (J_2^{\|}(k)-J_2^{\top}(0))-4\langle 
      S^y_B\rangle J_1^{\top}(0) \nonumber\\
&&M_{16}=4\langle S^y_A\rangle J_1^{\|}(k)\nonumber\\
&&M_{25}= 4\langle S^y_B\rangle J_1^{\|}(k)\nonumber\\
&&M_{26}=4\langle S^y_B\rangle (J_2^{\|}(k)-J_2^{\top}(0))-4\langle
      S^y_A\rangle J_1^{\top}(0)\nonumber\\ 
&&M_{35}=-4\langle S^x_A\rangle (J_2^{\|}(k)-J_2^{\top}(0))+4\langle
      S^x_B\rangle J_1^{\top}(0)\nonumber \\
&&M_{36}=-4\langle S^x_A\rangle J_1^{\|}(k)\nonumber\\
&&M_{45}=-4\langle S^x_B\rangle J_1^{\|}(k) \nonumber\\
&&M_{46}=-4\langle S^x_B\rangle (J_2^{\|}(k)-J_2^{\top}(0)) 
     +4\langle S^x_A\rangle J_1^{\top}(0)\nonumber\\
&&M_{51}= 4 \langle S^y_B\rangle J_1^{\top}(0)+4 \langle S^y_A \rangle
        (J_2^{\top}(0)-J_2^{\top}(k)) \nonumber\\
&&M_{52}=-4 \langle S^y_A\rangle J_1^{\top}(k)  \nonumber \\
&&M_{61}=-4\langle S^y_B\rangle J_1^{\top}(k) \nonumber\\
&&M_{62}= 4\langle S^y_A\rangle J_1^{\top}(0)+4\langle S^y_B \rangle 
       (J_2^{\top}(0)-J_2^{\top}(k))\nonumber \\
&&M_{53}= -4\langle S^x_B\rangle J_1^{\top}(0)-4\langle S^x_A\rangle
   (J_2^{\top}(0)-J_2^{\top}(k)) \nonumber\\
&&M_{54}= 4\langle S^x_A\rangle J_1^{\top}(k)\nonumber\\
&&M_{63}= 4\langle S^x_B\rangle J_1^{\top}(k) \nonumber \\
&&M_{64}= -4\langle S^x_A\rangle 
 J_1^{\top}(0)-4\langle S^x_B\rangle  (J_2^{\top}(0)-J_2^{\top}(k))\nonumber\\
\end{eqnarray}
Here the k-dependent coupling constants are defined by
$J_1^{\top}(k)=J_1^{\top}\gamma_1(k)$,
$J_2^{\top}(k)=J_2^{\top}\gamma_2(k)$,
$J_1^{\top}(k)=J_1^{\top}\gamma_1(k)$ and 
$J_2^{\top}(k)=J_2^{\top}\gamma_1(k)$,  where
\begin{eqnarray}
\gamma_1({\bf{k}})=\frac{1}{q_1} \sum_{a_{AB}} e^{i {\bf{k}}\,{\bf{a}}_{AB}}\nonumber\\
\gamma_2({\bf{k}})=\frac{1}{q_2} \sum_{a_{AA}} e^{i {\bf{k}}\,{\bf{a}}_{AA}}
\end{eqnarray}
are the lattice generating functions.
On the bcc lattice these lattice generating functions are given by
\begin{eqnarray}
\gamma_1({\bf{k}})=
\cos\left(\frac{k_x a}{2}\right)\cos\left(\frac{k_y a}{2}\right)
\cos\left(\frac{k_z a}{2}\right)\nonumber\\
\gamma_2({\bf{k}})=\frac{\cos\left(k_x a\right)}{3}+\frac{\cos\left(k_y a\right)}{3}+
\frac{\cos\left(k_z a\right)}{3}
\end{eqnarray}
where $a$ is the lattice constant of a simple cubic sub-lattice. 
Again, the spontaneously broken U(1)-symmetry 
gives us the freedom to set $\langle S^y_A\rangle =
\langle S^y_B\rangle =0$. This reduces 
the number of non-zero matrix components of $M$.

It has been shown that the commutator Green's functions 
must not have a zero frequency pole \cite{Nafari}.
This results directly from the fact that the commutator of to spin operators
at long time distances becomes zero:
$\lim_{t\rightarrow\infty} \langle [S^{\mu}(t) S^{\nu}] \rangle=0$.
In the present calculation the Green's functions do actually
acquire a zero frequency pole, as given by the eigenvalues of the matrix $M$.
In the ferromagnetic and anti-ferromagnetic phases 
this zero frequency pole is readily canceled out,
 but for the canted ferromagnetic and the canted anti-ferromagnetic phases this imposes an additional 
constraint:
\begin{eqnarray}\label{sssfcon}
h^z+2 \langle S^z_A\rangle (J_2^{\|}-J_2^{\top})+2 \langle S^z_B\rangle J_1^{\|}=
   2 J_1^{\top}\frac{\langle S^x_B\rangle }{\langle S^x_A\rangle }\langle S^z_A\rangle
   \nonumber\\
h^z+2\langle S^z_B\rangle (J_2^{\|}-J_2^{\top})+2\langle S^z_A\rangle J_1^{\|}=
   2 J_1^{\top}\frac{\langle S^x_A\rangle }{\langle S^x_B\rangle }\langle S^z_B \rangle\\
\end{eqnarray}
Note that these two conditions are identical
to the classical mean-field equations \cite{epl,epjb1,epjb2}.
We use these two relations to replace the external magnetic field $h_z$
in the matrix $M$. To combine terms we introduce following variables:
\begin{eqnarray}\label{supapara}
A_1= J_2^{\top}(1-\gamma_2(k)) 
    + J_1^{\top}\frac{\langle S^x_B\rangle }
                                  {\langle S^x_A\rangle }\nonumber\\
A_2= J_2^{\top}(1-\gamma_2(k)) 
    + J_1^{\top}\frac{\langle S^x_A\rangle }
                                  {\langle S^x_B\rangle }\nonumber\\
B_1= J_2^{\top}(1-\gamma_2(k)) 
    + J_1^{\top}\frac{\langle S^x_B\rangle }
                                  {\langle S^x_A\rangle }\nonumber\\
     -4(J_2^{\|}-J_2^{\top})\gamma_2(k)\langle S^x_A\rangle^2\nonumber\\
B_2= J_2^{\top}(1-\gamma_2(k)) 
    + J_1^{\top}\frac{\langle S^x_A\rangle }
                                   {\langle S^x_B \rangle }\nonumber\\
        -4(J_2^{\|}-J_2^{\top})\gamma_2(k)  \langle S^x_B\rangle^2 \nonumber \\
C= J_1^{\top}\gamma_1(k)\nonumber\\                                                  
D=-4\gamma_1(k)(J_1^{\top}\langle S^z_A\rangle\langle S^z_B\rangle +2
           J_1^{\|} \langle S^x_A\rangle
                   \langle S^x_B\rangle )
\end{eqnarray}
Then the matrix $M$ in the canted anti-ferromagnetic and canted ferromagnetic phases, 
where Equation (\ref{sssfcon}) holds,
is given by (in block-form):
\begin{eqnarray}
M_{\mathbf{c}}=\left(
\begin{array}{lll}
M^{\mathbf{c}}_{11}&M^{\mathbf{c}}_{12}&M^{\mathbf{c}}_{13}\\
M^{\mathbf{c}}_{21}&M^{\mathbf{c}}_{22}&M^{\mathbf{c}}_{23}\\
M^{\mathbf{c}}_{31}&M^{\mathbf{c}}_{32}&M^{\mathbf{c}}_{33}
\end{array}\right)
\end{eqnarray}
where the $2\times2$-blocks are given by: 
\begin{eqnarray}
&&M^{\mathbf{c}}_{11}=M^{\mathbf{c}}_{22}=M^{\mathbf{c}}_{33}=\left[
\begin{array}{ll}
i\omega &0\\
0&i\omega \\
\end{array}\right]\nonumber\\
&&M^{\mathbf{c}}_{12}=-M^{\mathbf{c}}_{21}=\left[
\begin{array}{ll}
  4 A_1\langle S^z_A\rangle  &  -4 C \langle S^z_A\rangle \\
    -4 C \langle S^z_B\rangle &  4 A_2\langle S^z_B\rangle  
\end{array}\right]\nonumber\\
&&M^{\mathbf{c}}_{13}=M^{\mathbf{c}}_{31}=\left[
\begin{array}{ll} 
0&0\\
0&0
\end{array}\right]\nonumber\\
&&M^{\mathbf{c}}_{23}=\left[
\begin{array}{ll}  
  - \frac{4 A_1\langle S^z_A\rangle^2-B_1}{\langle S^x_A\rangle}&
  \frac{4 C\langle S^z_A\rangle\langle S^z_B\rangle +D}
  {\langle S^x_B\rangle } \\ 
  \frac{4 C\langle S^z_A\rangle\langle S^z_B\rangle +D}{\langle S^x_A\rangle
   } &
  -\frac{4 A_2\langle S^z_B\rangle^2-B_2}{\langle S^x_B\rangle}
\end{array}\right]\nonumber\\
&&M^{\mathbf{c}}_{32}=\left[
\begin{array}{ll}
  4 -A_1\langle S^x_A\rangle  &  4 C \langle S^x_A\rangle \\
    4 C \langle S^x_B\rangle &  -4 A_2\langle S^x_B\rangle  
\end{array}\right]
\end{eqnarray}
In the phases that do not exhibit off-diagonal long range order, namely the 
ferromagnetic and anti-ferromagnetic phases, Equation (\ref{sssfcon}) 
does not hold. However, as $\langle S_x \rangle=\langle S_y \rangle=0$ 
the matrix $M$ reduced to a matrix of $4\times4$-dimensions:
\begin{eqnarray}
M^{\mathbf{nc}}=\left(
\begin{array}{rrrrrr}
 i\omega & 0    &M_{13}&M_{14} \\
 0     & i\omega&M_{23}&M_{24} \\ 
-M_{13}&-M_{14}&i\omega &0 \\
-M_{23}&-M_{24}&0&i\omega  \\
\end{array}
\right)
\end{eqnarray}
Equation (\ref{Meq}) readily implies  that the Green's functions are given by:
\begin{eqnarray}
M^{-1}\cdot v
\end{eqnarray}
As these Green's functions in explicit form are rather complex,
we defer presenting them in their full form until the end of this Chapter.
Those readers who are not interested in the details of these functions 
might safely skip the corresponding section of the Chapter.

The Green's functions derived above determine the state of the system
and all relevant macroscopic and thermodynamic properties
can be extracted from them.
Nevertheless the Green's functions as derived in this random phase approximation are function
not only of the external field $h^z$ but also of the mean fields
of the spins. Therefore we have to define self-consistency equation which determine
those mean-fields.

\section{Self-Consistency Equations at Zero Temperature}

The Green's functions as derived in the previous Chapter do not bear
an explicit dependence on the temperature but rather depend on the 
temperature through the various mean-fields. Thus, upon deriving the 
determining self-consistency equations the temperature will be introduced 
explicitly into the formalism. We have decided here to separate the zero temperature and 
the finite temperature formalism. Even though, it would generally be easy to carry 
out the limiting process $T\; \rightarrow 0$ at any time it would cause 
numerical difficulties to do so at a later stage as the temperature 
usually appears in terms of $1 / T$. Also we find it instructive to 
derive the zero temperature formalism separately as it provides unveiled 
insight into the quantum nature of the model.
 
Naturally there seem to be three ways to set up self-consistency equations. We 
could, for example calculate:
\begin{eqnarray}
\lefteqn{- \frac{1}{2}\langle  S^z_i \rangle =
 \langle i  [S^x_i(0),S^y_i] \rangle} \nonumber \\
&&=G^{xy}_{{ij}_{Adv}}(0)-G^{xy}_{{ij}_{Ret}}(0)\nonumber\\
&&=\int_{-\infty}^{\infty} d\omega [G^{xy}_{{ij}_{Adv}}
(\omega)-G^{xy}_{{ij}_{Ret}}(\omega)]
\end{eqnarray}
an equivalent relation holds for $\langle S^x_i \rangle$ and 
$G^{zy}_{ij}$. Here, for the sake of readability we have disregarded the sub-lattice
subscripts A and B. Unfortunately, these equations are not suitable 
to calculate the 
mean-fields self-consistently but rather lead to an identity,
 giving identically $\frac{i}{2}\langle S^z \rangle$ on both sides.
That is due to the structure of the equation of motion, and
the $\langle S^z_i \rangle$ are exactly those ones contained in the right hand side
of Equation (\ref{Meq}). Therefore the appropriate and only choice to
define self-consistency equations is:
\begin{eqnarray}
\langle  S^y(0)_i S^y_i  \rangle=\frac{1}{4}
\end{eqnarray}  
In  order to establish the link between the correlation 
functions and the corresponding Green's function 
the following spectral expansion of the Green's function at absolute zero
as is readily obtained from Equation (\ref{GFdef}):
\begin{eqnarray}\label{specrep}
G^{yy}_{{ij}_{Ret/Adv}}(\omega)&=&
\sum_{n} \frac{\langle n_0| S^y_i|n\rangle\langle n| S^y_j |n_0 \rangle}
         {\omega-(\omega_n-\omega_{n_0})\pm i\epsilon}\nonumber \\
&&-\sum_{n} \frac{\langle n_0| S^y_j |n\rangle\langle n|  S^y_i  |n_0 \rangle}
         {\omega+(\omega_n-\omega_{n_0})\pm i\epsilon}
\end{eqnarray}
where we have used that:
\begin{eqnarray}
\frac{1}{2 \pi} \int_{-\infty}^{\infty} e^{i(\omega-\omega')t}
\theta (t)dt=\lim_{\epsilon\rightarrow +0}
\frac{1}{2\pi}\frac{i}{\omega-\omega'+i\epsilon} 
\end{eqnarray}
Here $|n_o \rangle$ is the ground state of the system and $| n \rangle$
refers to the complete set of eigenstates.
The spectral representation (Equation (\ref{specrep}))  shows
that the integral:
\begin{figure}[t]
\centering
\includegraphics[width=13cm]{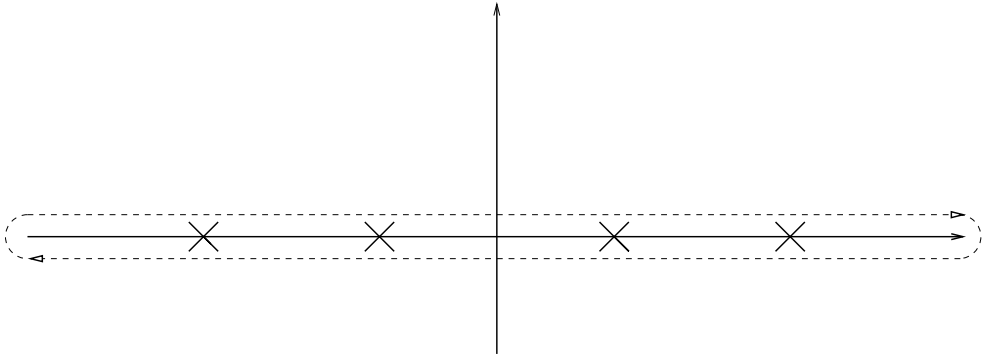} 
\caption{\label{fig:contplot} The correlation functions are given by 
contour integrals which circumscribe the four poles. The residue theorem
states that the integral equals the  sum over the four residues.}
\end{figure}
\begin{eqnarray}
\lefteqn{\int_{-\infty}^{\infty} d\omega e^{i \omega t} (G^{yy}_{{ij}_{Adv}}
(\omega)-G^{yy}_{{ij}_{Ret}}(\omega))}\nonumber \\&&
=\lim_{\epsilon\rightarrow 0} \int_{-\infty}^{\infty} d\omega
e^{i\omega t} (G^{yy}_{{ij}}
(\omega+i\epsilon)-G^{yy}_{{ij}}(\omega-i\epsilon))
\end{eqnarray}
is a contour integral enclosing the poles
of the Fourier transformed Green's function, as
can be seen in Figure (\ref{fig:contplot}).
The path in the upper and lower half planes correspond to the 
retarded and Green's functions respectively.
According to the residue theorem of complex analysis 
the value of the integral is given by the sum of the 
residues of the enclosed poles.
As $\omega_{n_0}$ is by definition the lowest energy state,
the spectral representation
of the Green's function shows that the correlation
function $\langle S^y_i (t)S^y_j \rangle$ corresponds
 to all negative poles and and the conjugated correlations function
$\langle S^y_j S^y_i (t) \rangle$ corresponds
to all positive poles. Hence we obtain the  desired
correlation function $\langle S^y_i (t)S^y_j \rangle$ 
by restricting the contour integral to 
negative frequencies:
\begin{eqnarray} \label{rcorrf}
\lefteqn{
\langle S^y_{i}(t)S^y_{j}\rangle=}\nonumber\\
&&-i \lim_{\epsilon\rightarrow 0}
                   \int_{-\infty}^{0}  e^{i\omega t}
                    G^{yy}_{ij}(\omega+i\epsilon)-
                         G^{yy}_{ij}(\omega-i\epsilon)
                         d\omega \nonumber\\
&&+\langle n_0| S^y_i|n_0\rangle\langle n_0| S^y_j |n_0 \rangle
\end{eqnarray}
The second term on the right side appears because its
contribution in the spectral representation (Equation (\ref{specrep}))
is canceled out.
As we have broken the 
U(1) symmetry of the ground state in a way that $\langle S^y \rangle=0$
this term will yield zero anyway and we will disregard it in further 
discussion.
Using the Green's function's Fourier transform into k-space
we obtain:
\begin{eqnarray}
\lefteqn{
\langle S^y_{i}(t)S^y_{j}\rangle=}\nonumber \\
&& -i \lim_{\epsilon\rightarrow 0}
\int\int_{-\infty}^{0} d^3k\, d\omega e^{i(k r_{ij}+\omega t)}
                    [G^{yy}(k,\omega+i\epsilon)\nonumber \\
&&  \quad                 -G^{yy}(k,\omega-i\epsilon)]
                          d\omega\nonumber\\
&&=2 \pi \int d^3k\, e^{ikr_{ij}} 
    [\mbox{Residue}\left(G^{yy}(k,\omega)e^{i\omega t},
                    -\omega_1\right)\nonumber \\
&&    \quad       +\mbox{Residue}\left(G^{yy}(k,\omega)e^{i\omega t},
                    -\omega_2\right) ]\nonumber\\
\end{eqnarray}
Now we can define two self-consistency equation which determine the 
spin fields:
\begin{eqnarray}\label{eq:selfonst0k}
\lefteqn{F_A:=2\pi\int d^3k\, 
                    [\mbox{Residue}
                    (G^{yy}_{A}(k,\omega),-\omega_1)}\nonumber \\
  &&                       +\mbox{Residue}(G^{yy}_{A}(k,\omega),-\omega_2)]
                        -\frac{1}{4}=0 \nonumber\\
\lefteqn{F_B:=2\pi\int d^3k \,
                  [\mbox{Residue}
                    (G^{yy}_{B}(k,\omega),-\omega_1)}\nonumber \\
  &&                       +\mbox{Residue}(G^{yy}_{B}(k,\omega),-\omega_2)]
                        -\frac{1}{4}=0  
 \end{eqnarray}
In the canted anti-ferromagnetic and canted ferromagnetic phases we use 
Equation (\ref{sssfcon}) to eliminate $\langle S^z_A \rangle$ and 
$\langle S^z_B \rangle$ and re-introduce $h^z$ instead and obtain:
\begin{eqnarray}\label{scszt}
F_A^c(\langle S^x_A\rangle ,\langle S^x_B\rangle,h^z)=0 \nonumber\\
F_B^c(\langle S^x_A\rangle ,\langle S^x_B\rangle,h^z)=0
\end{eqnarray}
similarly, we obtain the self-consistency equations for the ferromagnetic
and anti-ferromagnetic phases:
\begin{eqnarray}\label{scnzt}
 F_A^{nc}(\langle S^z_A\rangle ,\langle S^z_B\rangle,h^z)=0 \nonumber\\
  F_B^{nc}(\langle S^z_A\rangle ,\langle S^z_B\rangle,h^z)=0
 \end{eqnarray}
the three dimensional numerical integral contained in those equations
can be reduced by one dimension by introducing a two dimensional density 
of states $\rho(\gamma_1',\gamma_2')=\int d^3k \delta(\gamma_1'-\gamma_1(k))\delta(\gamma_2'-\gamma_2(k))$,
so that:
\begin{eqnarray}\label{eq:dos}
\int d^3k \quad\rightarrow\quad\int d\gamma_1d\gamma_2 \;\rho(\gamma_1,\gamma_2)
\end{eqnarray}
Consequently, in the self-consistency equations the DOS  $\rho(\gamma_1,\gamma_2)$ 
is the only remaining term depending on the specific lattice structure.
(In the actual numerical calculation it is, in order to avoid singularities in the origin,
feasible to integrate over $\rho(\arccos(\gamma_1),\arccos(\gamma_2))$.For small $k$, 
$\arccos(\gamma_1)$ and $\arccos(\gamma_2))$ vary as $\sim k$ and therefore take
the important contributions at small k more accurately into account than 
$\gamma_1$ and $\gamma_2$ which vary as $\sim k^2$.)
Through this procedure we have archived a wider applicability
of the corresponding equations as other systems, i.e. lattice geometries exhibiting frustration or
two dimensional systems where linear spin-waves are still a valid approximation, can easily be 
accomplished by simply inserting the appropriate DOS. The DOS for the bipartite bcc 
lattice is shown in Figure \ref{fig:dos}.
\begin{figure}[t]
\centering
\includegraphics[width=13cm]{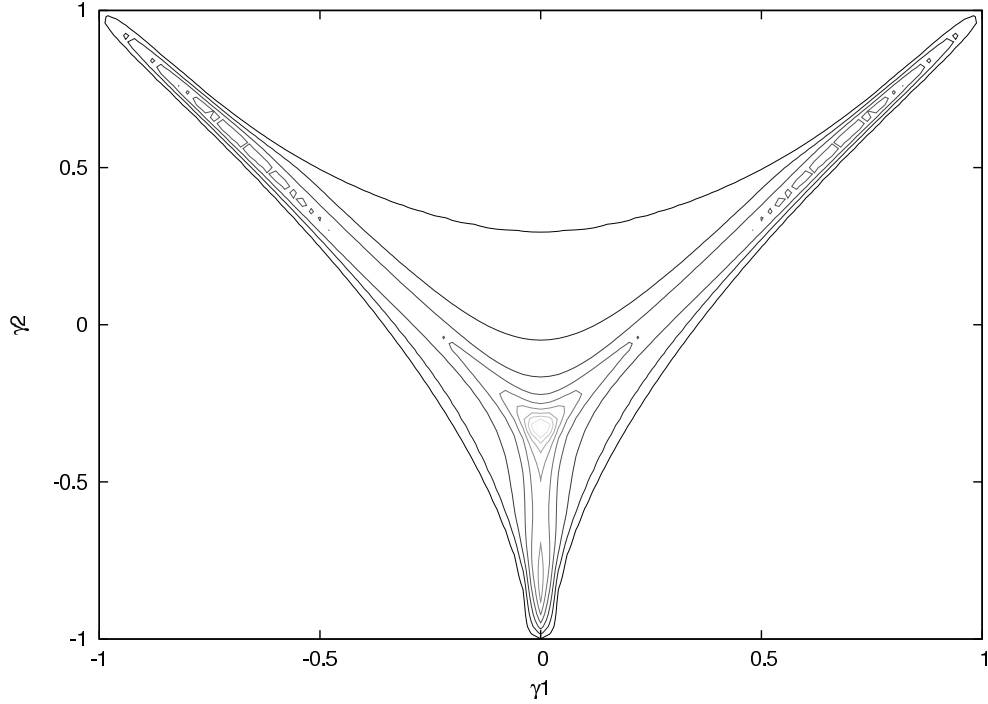} 
\caption{\label{fig:dos} Two dimensional density of state $\rho(\gamma_1,\gamma_2)$ 
for the bipartite bcc lattice. The invariance of the DOS under $\gamma_1\;\rightarrow\;-\gamma_1$ 
reflects the discrete translational symmetry of the bcc lattice along the diagonal, i.e. $\{x,y,z\}\;
\rightarrow\;\{x+a/2,y+a/2,z+a/2\}$. The main contribution to the density of states comes from a relatively small area
around three lines in the shape of the letter Y with a peak at $\gamma_1=0$ and $\gamma_2=-0.327$.}
\end{figure}

\section{Self-Consistency Equations at Finite Temperature}
\label{sec:selfconftf}

At finite temperature the system additionally exhibits thermal fluctuations 
expressed thorough averaging with the Boltzmann weight. As a consequence
the negative and positive poles of the Green's functions do no longer refer solely to the 
corresponding correlations functions or their conjugates respectively. As the temperature is 
turned on the contribution of the poles to the correlation functions starts to overlap with the 
contribution of their conjugate correlation functions, where the degree of overlap is determined by the 
Boltzmann weight. Therefore the extraction of the correlations function from the corresponding
commutator Green's function is not as straight forward as in the zero temperature case but  
the overlap  to be reversed, i.e. an additional factor cancels out the contribution of the conjugate function.
 This was done by Tyablikov \cite{Bogo,Tyablikov} in 1959. Here we follow a slightly different derivation.

As we have seen in the disquisition on the 
zero-temperature regime the adequate self-consistency equations
are expressed by:
\begin{eqnarray}
\langle S^y(0)_i S^y_i \rangle=\frac{1}{4}
\end{eqnarray}  
The spectral representation of the Green's function at finite temperature 
is according to Equation (\ref{GFdef}) given by:
\begin{eqnarray}\label{specrepft}
\lefteqn{G^{yy}_{{ij}_{Ret/Adv}}(\omega)=}\nonumber\\
&&\phantom{-}\sum_{m,n}\frac{1}{2 \pi} \frac{\langle n| S^y_i|m\rangle\langle m| S^y_j |n \rangle}
         {\omega-(\omega_m-\omega_{n})\pm i\epsilon}\frac{e^{-\beta \omega _n}}{Z}\nonumber \\
&&-\sum_{m,n} \frac{1}{2 \pi}\frac{\langle n| S^y_j |m\rangle\langle m|  S^y_i  |n \rangle}
         {\omega+(\omega_m-\omega_{n})\pm i\epsilon}\frac{e^{-\beta \omega _n}}{Z}\nonumber\\
&&=
\sum_{m,n}\frac{1}{2 \pi} \frac{\langle n| S^y_i|m\rangle\langle m| S^y_j |n \rangle}
         {\omega-(\omega_m-\omega_{n})\pm i\epsilon}\frac{e^{-\beta \omega _n}-e^{-\beta \omega _m}}{Z}
\end{eqnarray}
where we once again used that:
\begin{eqnarray}
\frac{1}{2 \pi} \int_{-\infty}^{\infty} e^{i(\omega-\omega')t}
\theta (t)dt=\lim_{\epsilon\rightarrow +0}
\frac{1}{2\pi}\frac{i}{\omega-\omega'+i\epsilon} 
\end{eqnarray}
$|m \rangle$ and $|n \rangle$ are eigenstates of the Hamiltonian and $Z$
refers to the partition function. 

The Fourier transform of the correlation function is given by:
\begin{eqnarray}
\langle S^y_{i}(t)S^y_{j}\rangle_{\omega}&=&
\sum_{m,n} \langle n| S^y_i|m\rangle\langle m| S^y_j |n \rangle
         \frac{e^{-\beta \omega_n}}{Z}\nonumber\\
&&\times \delta(\omega-(\omega_m-\omega_{n})) 
\end{eqnarray}
The relation to the Green's function is readily obtained by the
following rearrangements:
\begin{eqnarray}
\lefteqn{\langle S^y_{i}(t)S^y_{j}\rangle_{\omega}=
\sum_{m,n `}\frac{ \langle n| S^y_i|m\rangle\langle m| S^y_j |n \rangle}{Z}
        e^{-\beta \omega_n}}\nonumber\\
&& \phantom{=}  \times\frac{e^{\beta(\omega_n-\omega_m)}-1}{e^{\beta(\omega_n-\omega_m)}-1}
 \delta(\omega-(\omega_m-\omega_{n})) \nonumber \\
&&\phantom{=} +\sum_{n}\frac{ \langle n| S^y_i|n\rangle\langle n| S^y_j |n \rangle}{Z} e^{-\beta \omega_n}\nonumber\\
&&=\frac{\sum_{m,n `} \langle n| S^y_i|m\rangle\langle m| S^y_j |n \rangle}{e^{\beta \omega}-1}
        \frac{e^{-\beta \omega_m}-e^{-\beta \omega_n}}{Z}\nonumber\\
&&\phantom{=} \times \delta(\omega-(\omega_m-\omega_{n})) \nonumber \\
&&\phantom{=} +\sum_{n}\frac{ \langle n| S^y_i|n\rangle\langle n| S^y_j |n \rangle}{Z} e^{-\beta \omega_n}\nonumber\\
\end{eqnarray}
The prime ' on the sum excludes terms where $m=n$ and 
was introduced to avoid singularities in the denominator.
The omitted terms are accounted for in the second term of the RHS.
The following symbolic identity is valid for real $\omega$
\begin{eqnarray}
\lim_{\epsilon\rightarrow +0} 
\left[\frac{1}{\omega-\omega'+i\epsilon}
     -\frac{1}{\omega-\omega'-i\epsilon} \right]
=-2\pi i\delta(\omega-\omega')
\end{eqnarray} 
and immediately yields:
\begin{eqnarray} \label{rcorrfft}
\lefteqn{
\langle S^y_{i}(t)S^y_{j}\rangle=}\nonumber\\
&&i \lim_{\epsilon\rightarrow 0}
                   \int_{-\infty}^{0}  e^{i\omega t}
                    \frac{G^{yy}_{ij}(\omega+i\epsilon)-
                         G^{yy}_{ij}(\omega-i\epsilon)
                         }{e^{\beta\omega}-1}d\omega \nonumber\\
&& +\sum_{n}\frac{ \langle n| S^y_i|n\rangle\langle n| S^y_j |n \rangle}{Z} e^{-\beta \omega_n}
\end{eqnarray}
As pointed out by Stevens and Toombs \cite{Stevens} the second term on the right side is generally 
difficult to calculate. In principle it would have been possible to eliminate this term 
by also calculating the anti-commutator Green's function. Once again, as we have broken the
U(1) symmetry of the ground state in a way such that $\langle S^y \rangle=0$ this term 
will yield zero.

Using the Green's function's Fourier transform into k-space
we obtain:
\begin{eqnarray}
\lefteqn{
\langle S^y_{i}(t)S^y_{j}\rangle=}\nonumber \\
&& i\lim_{\epsilon\rightarrow 0}
\int\int_{-\infty}^{0} d^3k\,d\omega\,\frac{ e^{i(k r_{ij}+\omega t)}}{e^{\beta \omega}-1}\nonumber\\
&&                    [G^{yy}(k,\omega+i\epsilon)
                  -G^{yy}(k,\omega-i\epsilon)]
                          d\omega\nonumber\\
&&=-2 \pi  \int d^3k \, e^{ikr_{ij}}\sum_{\omega_p=\omega_1,\omega_2} 
    [\mbox{ Residue}\left(\frac{G^{yy}(k,\omega)e^{i\omega t}}{e^{\beta\omega}-1},
                    \omega_p\right)\nonumber \\
&&    \quad       +\mbox{Residue}\left(\frac{G^{yy}(k,\omega)e^{i\omega t}}{e^{\beta\omega}-1},
                    -\omega_p\right) ]\nonumber\\
\end{eqnarray}
Now we can define two self-consistency equations which determine the 
spin fields at finite temperature:
\begin{eqnarray}\label{eq:selfconftfk}
\lefteqn{F_A:=\int d^3k\, \sum_{\omega_p}
                    [\mbox{Residue}
                    (\frac{G^{yy}_{A}(k,\omega)}{e^{\beta \omega}-1},\omega_p)}\nonumber \\
  &&                       +\mbox{Residue}(\frac{G^{yy}_{A}(k,\omega)}{e^{\beta \omega}-1},-\omega_p)]
                        -\frac{1}{4}=0 \nonumber\\
\lefteqn{F_B:=\int d^3k \,\sum_{\omega_p}
                  [\mbox{Residue}
                    (\frac{G^{yy}_{B}(k,\omega)}{e^{\beta \omega}-1},\omega_p)}\nonumber \\
  &&                       +\mbox{Residue}(\frac{G^{yy}_{B}(k,\omega)}{e^{\beta \omega}-1},-\omega_p)]
                        -\frac{1}{4}=0  
\end{eqnarray}
Using the DOS of Equation (\ref{eq:dos}) those are written as:
\begin{eqnarray}\label{eq:selfconftfg}
\lefteqn{F_A=\int d\gamma_1 \,d\gamma_2\, \rho(\gamma_1,\gamma_2) \times}\nonumber\\&&
		  \sum_{\omega_p} [\mbox{Residue}
                     (\frac{G^{yy}_{A}(\gamma_1,\gamma_2,\omega)}{e^{\beta \omega}-1},\omega_p)\nonumber \\
  &&                       +\mbox{Residue}(\frac{G^{yy}_{A}(\gamma_1,\gamma_2,\omega)}{e^{\beta \omega}-1},-\omega_p)]
                        -\frac{1}{4}=0 \nonumber\\
\lefteqn{F_B=\int d\gamma_1 \,d\gamma_2\, \rho(\gamma_1,\gamma_2)\times}\nonumber\\ &&
                 \sum_{\omega_p} [\mbox{Residue}
                    (\frac{G^{yy}_{B}(\gamma_1,\gamma_2,\omega)}{e^{\beta \omega}-1},\omega_p)\nonumber \\
  &&                       +\mbox{Residue}(\frac{G^{yy}_{B}(\gamma_1,\gamma_2,\omega)}{e^{\beta \omega}-1},-\omega_p)]
                        -\frac{1}{4}=0  
\end{eqnarray}
Again, in the canted anti-ferromagnetic and canted ferromagnetic phases 
we use the mean-field equations \cite{epl,epjb1,epjb2} to eliminate 
$\langle S^z_A \rangle$ 
and $\langle S^z_B \rangle$ and obtain:
\begin{eqnarray}\label{scs}
F_A^c(\langle S^x_A\rangle ,\langle S^x_B\rangle,h^z,T)=0 \nonumber\\
F_B^c(\langle S^x_A\rangle ,\langle S^x_B\rangle,h^z,T)=0
\end{eqnarray}
Similarly, we obtain the self-consistency equations for the ferromagnetic
and anti-ferromagnetic phases:
\begin{eqnarray}\label{scn}
 F_A^{nc}(\langle S^z_A\rangle ,\langle S^z_B\rangle,h^z,T)=0 \nonumber\\
  F_B^{nc}(\langle S^z_A\rangle ,\langle S^z_B\rangle,h^z,T)=0
 \end{eqnarray}

\section{Green's Functions in Explicit Form}

In the remaining three Chapters we give the Green's functions and the corresponding 
correlations functions and carry out the derivation of the matrix equation defining the 
Green's functions. 

We start with the Green's functions for the anisotropic Heisenberg model 
for the canted phases, i.e. for the canted ferromagnetic and the canted
anti-ferromagnetic phases
\begin{eqnarray}
  \lefteqn{ G_{A}^{xy}(k,\omega)=
   \frac{-i  \omega\langle S^z_A\rangle }{(2\pi)^4( \omega ^2-\omega_1^2) 
   ( \omega ^2-\omega_2^2)}
   \times} \nonumber\\
&&   ( \omega ^2+4 C D+4 D A_1 
   -4 (C+A_2) (B_2+A_2 (4
   \langle S^x_B\rangle^2+4\langle S^z_B\rangle^2-1)))\nonumber
\end{eqnarray}
\begin{eqnarray}
 \lefteqn{G_{B}^{xy}(k,\omega)=
   \frac{-i  \omega\langle S^z_B\rangle }{(2\pi)^4( \omega ^2-\omega_1^2) 
   ( \omega ^2-\omega_2^2)}
   \times}\nonumber\\
&&   ( \omega ^2+4 C D+4 D A_2-
   4 (C+A_1) (B_1+A_1 (4
  \langle S^x_A\rangle^2+4\langle S^z_A\rangle^2-1)))
\end{eqnarray}
\begin{eqnarray}
  \lefteqn{ G_{A}^{yy}(k,\omega)=
  \frac{-1 }{(2\pi)^4( \omega ^2-\omega_1^2) 
   ( \omega ^2-\omega_2^2)}
   \times}\nonumber\\
&&(D ( \omega ^2+4 C D+4 D A_2)+(B_1+A_1 (4\langle S^x_A\rangle^2+4
  \langle S^z_A\rangle^2-1))\nonumber\\
&&  ( \omega ^2-4(C+A_2) (B_2+A_2 (4\langle S^x_B\rangle^2+4
  \langle S^z_B\rangle^2-1))))\nonumber
\end{eqnarray}
\begin{eqnarray}
 \lefteqn{   G_{B}^{yy}(k,\omega)=
  \frac{-1 }{(2\pi)^4( \omega ^2-\omega_1^2) 
   ( \omega ^2-\omega_2^2)}
   \times}\nonumber\\
&&(D ( \omega ^2+4 C D+4 D A_1)+(B_1+A_1 (4\langle S^x_B\rangle^2+4
  \langle S^z_B\rangle^2-1)) \nonumber\\
&&( \omega ^2-4(C+A_1) (B_1+A_1 (4\langle S^x_A\rangle^2+4
  \langle S^z_A\rangle^2-1))))
\end{eqnarray}
\begin{eqnarray}
 \lefteqn{ G_{A}^{zy}(k,\omega)=
   \frac{-i  \omega\langle S^x_A\rangle}{(2\pi)^4( \omega ^2-\omega_1^2) 
   ( \omega ^2-\omega_2^2)}
   \times} \nonumber\\
&& ( \omega ^2+4 C D+4 D A_1-
   4 (C+A_2) (B_2+A_2 (4
  \langle S^x_B\rangle^2+4\langle S^z_B\rangle^2-1)))
\nonumber
\end{eqnarray}
\begin{eqnarray}
  \lefteqn{ G_{B}^{zy}(k,\omega)=
   \frac{-i  \omega\langle S^x_B\rangle}{(2\pi)^4( \omega ^2-\omega_1^2) 
   ( \omega ^2-\omega_2^2)}
   \times}\nonumber\\
&& ( \omega ^2+4 C D+4 D A_2-
   4 (C+A_1) (B_1+A_1 (4
   \langle S^x_A\rangle^2+4\langle S^z_A\rangle^2-1)))
\end{eqnarray}
The poles are given by:
\begin{eqnarray}
\lefteqn{ \omega_{1/2}=
2 (A_1 (B_1+A_1 (4\langle S^x_A\rangle^2+4\langle S^z_A\rangle^2-1))+}\nonumber\\&&
A_2 (B_2+A_2 (4\langle S^x_B\rangle^2+4
 \langle  S^z_B\rangle^2-1)))-4 C D\nonumber\\&&
   \pm
   2 [(-2 C D+A_1 (B_1+A_1 (4\langle S^x_A\rangle^2+4\langle S^z_A\rangle^2-1))\nonumber\\&&
+A_2(B_2+A_2 (4\langle S^x_B\rangle^2+4\langle S^z_B\rangle^2-1))){}^2\nonumber\\&&
-4 (C^2-A_1 A_2)
   (D^2-(B_1+A_1(4\langle S^x_A\rangle^2+4\langle S^z_A\rangle^2-1)) \nonumber\\&&
   \times 
  (B_2+A_2 (4\langle S^x_B\rangle^2+4\langle S^z_B\rangle^2-1)))]^{1/2}
\end{eqnarray}
All parameter $A_1$,$A_2$,$B_1$,$B_2$, $C$ and $D$ are as stated in 
Equation (\ref{supapara}).

As condition (\ref{sssfcon}) does not apply for the ferromagnetic 
nor for the anti-ferromagnetic states, these Green's functions are 
structurally different:
\begin{eqnarray}
   G_{A}^{xy}(k,\omega)=
\frac{i \omega  ((A_3+A_4) B_3\langle S^z_B\rangle-(- ^2\omega ^2 +A_4^2+B_3 B_4)
  \langle S^z_A\rangle)}{(2\pi)^4( ^2 \omega ^2- ^2 \omega _1^2) ( ^2 \omega ^2- ^2 \omega _2^2)}
\nonumber
\end{eqnarray}
\begin{eqnarray}
   G_{B}^{xy}(k,\omega)=
\frac{i \omega  ((A_3+A_4) B_4\langle S^z_A\rangle-(- ^2 \omega ^2+A_3^2+B_3 B_4)
  \langle S^z_B\rangle)}{(2\pi)^4 ( ^2 \omega ^2- ^2 \omega _1^2) ( ^2 \omega ^2- ^2 \omega _2^2)}
\end{eqnarray}
\begin{eqnarray}
  \lefteqn{ G_{A}^{yy}(k,\omega)=
\frac{1}
   {(2\pi)^4(\omega^2-\omega _1^2) 
   (\omega^2-\omega _2^2)} \times}\nonumber\\
&&((\omega^2 A_3 -A_3 A_4^2+A_4 B_3 B_4)
   \langle S^z_A\rangle
+B_3 (\omega^2 +A_3 A_4-B_3 B_4)\langle S^z_B\rangle)
\nonumber
\end{eqnarray}
\begin{eqnarray}
  \lefteqn{ G_{B}^{yy}(k,\omega)=
\frac{1}{(2\pi)^4 (\omega ^2-\omega _1^2) (\omega ^2-\omega _2^2)}\times}\nonumber\\
&&
(B_4 ( ^2\omega ^2 +A_3 A_4-B_3 B_4)\langle S^z_A\rangle+
( ^2\omega ^2 A_4 -A_3^2 A_4+A_3
   B_3 B_4)\langle S^z_B\rangle)
\end{eqnarray}
with
\begin{eqnarray}
\omega_{1/2}=\frac{A_3+A_4\pm \sqrt{(A_3-A_4)^2+4 B_3 B_4}}{2}
\end{eqnarray}
\begin{eqnarray}
A_3=2h_z+4\langle S^z_A\rangle (J_2^{\|}-J_2^{\top} \gamma_2(k))+4\langle S^z_B\rangle J^{\|}_1 \nonumber\\
A_4=2h_z+4\langle S^z_B\rangle (J_2^{\|}-J_2^{\top} \gamma_2(k))+4\langle S^z_A\rangle J^{\|}_1 \nonumber\\
B_3=-4\langle S^z_A\rangle J_1^{\top}\gamma_1 (k)
\nonumber\\
B_4=-4\langle S^z_B\rangle J_1^{\top}\gamma_1 (k)
\end{eqnarray}

\subsection{The Correlation Functions in Explicit Form}

Here we state the relevant correlation functions. All other correlation functions
yield identically $\frac{1}{4} \langle S_{x_{A/B}} \rangle$ or $\frac{1}{4} \langle S_{z_{A/B}} \rangle$
and are therefore not suitable as self-consistency equations.
We start with the Green's function for the canted anti-ferromagnetic and the 
canted ferromagnetic phase at finite temperature:
\begin{eqnarray}
\lefteqn{\langle S^y_i S^y_{j\in A}(0) \rangle_k=
-\frac{\coth\left(\frac{\beta \omega_1}{2}\right)}{2
   (\omega_1^3-\omega_1 \omega_2^2)}\times} \nonumber\\&&
  \Big(4 CD^2+\omega_1^2
   D+\omega_1^2 B_1-4 C B_1 B_2 \nonumber\\&&
   -4 A_2^2 B_1 (4\langle S^x_B\rangle^2+4\langle S^z_B\rangle^2-1)\nonumber\\&&
  + A_1(4\langle S^x_A\rangle^2+4\langle S^z_A\rangle^2-1)\times\nonumber\\&&
 \big(\omega_1^2-4 C B_2-
   4 A_2^2 (4\langle S^x_B\rangle^2+4\langle S^z_B\rangle^2-1)\nonumber\\&&
   -4 A_2 (B_2+C (4\langle S^x_B\rangle^2+4\langle S^z_B\rangle^2-1))\big)\nonumber\\&&
   +4 A_2\big(D^2-B_1 (B_2+C (4\langle S^x_B\rangle^2+4\langle S^z_B\rangle^2-1))\big)\Big)\nonumber\\&&
   -\frac{\coth\left(\frac{\beta \omega_2}{2}\right)}{2 (\omega_2^3-\omega_1^2 \omega_2)} 
   \Big(4 C D^2+\omega_2^2 D+\omega_2^2 B_1-4 C B_1 B_2\nonumber\\&&
   -4 A_2^2 B_1 (4\langle S^x_B\rangle^2+4\langle S^z_B\rangle^2-1)
   \nonumber\\&&
  + A_1 (4\langle S^x_A\rangle^2+4\langle S^z_A\rangle^2-1)\times\nonumber\\&&
 \big(\omega_2^2-4 C B_2-
   4 A_2^2 (4\langle S^x_B\rangle^2+4\langle S^z_B\rangle^2-1)\nonumber\\&&
   -4A_2 (B_2+C (4\langle S^x_B\rangle^2+4\langle S^z_B\rangle^2-1))\big)\nonumber\\&&
   +4 A_2 \big(D^2-B_1(B_2+C (4\langle S^x_B\rangle^2+4\langle S^z_B\rangle^2-1))\big)\Big)
\end{eqnarray}
\begin{eqnarray}
\lefteqn{\langle S^y_i S^y_{j\in B}(0) \rangle_k=
-\frac{\coth\left(\frac{\beta \omega_1}{2}\right)}{2
   (\omega_1^3-\omega_1 \omega_2^2)}\times}\nonumber\\&&
  \Big(4 CD^2+\omega_1^2
   D+\omega_1^2 B_2-4 C B_2 B_1 \nonumber\\&&
   -4 A_1^2 B_2 (4\langle S^x_A\rangle^2+4\langle S^z_A\rangle^2-1)\nonumber\\&&
   +A_2(4\langle S^x_B\rangle^2+4\langle S^z_B\rangle^2-1)\times\nonumber\\&& 
   \big(\omega_1^2-4 C B_1-
   4 A_1^2 (4\langle S^x_A\rangle^2+4\langle S^z_A\rangle^2-1)\nonumber\\&&
   -4 A_1 (B_1+C (4\langle S^x_A\rangle^2+4\langle S^z_A\rangle^2-1))\big)+\nonumber\\&&
   4 A_1\big(D^2-B_2 (B_1+C (4\langle S^x_A\rangle^2+4\langle S^z_A\rangle^2-1))\big)\Big)\nonumber\\&&
   -\frac{\coth\left(\frac{\beta \omega_2}{2}\right)}{2 (\omega_2^3-\omega_1^2 \omega_2)} 
   \Big(4 C D^2+\omega_2^2 D+\omega_2^2 B_2-4 C B_2 B_1\nonumber\\&&
   -4 A_2^2 B_2 (4\langle S^x_A\rangle^2+4\langle S^z_A\rangle^2-1)\nonumber\\&&
   +A_1 (4\langle S^x_B\rangle^2+4\langle S^z_B\rangle^2-1)\times\nonumber\\&&
    \big(\omega_2^2-4 C B_1-
   4 A_1^2 (4\langle S^x_A\rangle^2+4\langle S^z_A\rangle^2-1)\nonumber\\&&
   -4A_1 (B_1+C (4\langle S^x_A\rangle^2+4\langle S^z_A\rangle^2-1))\big)\nonumber\\&&
   +4 A_1 \big(D^2-B_2(B_1+C (4\langle S^x_A\rangle^2+4\langle S^z_A\rangle^2-1))\big)\Big)
\end{eqnarray}
At zero temperature these functions become:
\begin{eqnarray}
\lefteqn{\langle S^y_i S^y_{j\in A}(0) \rangle_k=
-\frac{1}{2
   \sqrt{\omega_1^2}(\omega_1^2- \omega_2^2)}\times} \nonumber\\&&
  \Big(4 CD^2+\omega_1^2
   D+\omega_1^2 B_1-4 C B_1 B_2 \nonumber\\&&
   -4 A_2^2 B_1 (4\langle S^x_B\rangle^2+4\langle S^z_B\rangle^2-1)\nonumber\\&&
  + A_1(4\langle S^x_A\rangle^2+4\langle S^z_A\rangle^2-1)\times\nonumber\\&&
 \big(\omega_1^2-4 C B_2-
   4 A_2^2 (4\langle S^x_B\rangle^2+4\langle S^z_B\rangle^2-1)\nonumber\\&&
   -4 A_2 (B_2+C (4\langle S^x_B\rangle^2+4\langle S^z_B\rangle^2-1))\big)\nonumber\\&&
   +4 A_2\big(D^2-B_1 (B_2+C (4\langle S^x_B\rangle^2+4\langle S^z_B\rangle^2-1))\big)\Big)\nonumber\\&&
   -\frac{1}{2\sqrt{\omega_2^2} (\omega_2^2-\omega_1^2 )} 
   \Big(4 C D^2+\omega_2^2 D+\omega_2^2 B_1-4 C B_1 B_2\nonumber\\&&
   -4 A_2^2 B_1 (4\langle S^x_B\rangle^2+4\langle S^z_B\rangle^2-1)
   \nonumber\\&&
  + A_1 (4\langle S^x_A\rangle^2+4\langle S^z_A\rangle^2-1)\times\nonumber\\&&
 \big(\omega_2^2-4 C B_2-
   4 A_2^2 (4\langle S^x_B\rangle^2+4\langle S^z_B\rangle^2-1)\nonumber\\&&
   -4A_2 (B_2+C (4\langle S^x_B\rangle^2+4\langle S^z_B\rangle^2-1))\big)\nonumber\\&&
   +4 A_2 \big(D^2-B_1(B_2+C (4\langle S^x_B\rangle^2+4\langle S^z_B\rangle^2-1))\big)\Big)
\end{eqnarray}
\begin{eqnarray}
\lefteqn{\langle S^y_i S^y_{j\in B}(0) \rangle_k=
-\frac{1 }{2 \sqrt{\omega_1^2}
   (\omega_1^2- \omega_2^2)}\times}\nonumber\\&&
  \Big(4 CD^2+\omega_1^2
   D+\omega_1^2 B_2-4 C B_2 B_1 \nonumber\\&&
   -4 A_1^2 B_2 (4\langle S^x_A\rangle^2+4\langle S^z_A\rangle^2-1)\nonumber\\&&
   +A_2(4\langle S^x_B\rangle^2+4\langle S^z_B\rangle^2-1)\times\nonumber\\&& 
   \big(\omega_1^2-4 C B_1-
   4 A_1^2 (4\langle S^x_A\rangle^2+4\langle S^z_A\rangle^2-1)\nonumber\\&&
   -4 A_1 (B_1+C (4\langle S^x_A\rangle^2+4\langle S^z_A\rangle^2-1))\big)+\nonumber\\&&
   4 A_1\big(D^2-B_2 (B_1+C (4\langle S^x_A\rangle^2+4\langle S^z_A\rangle^2-1))\big)\Big)\nonumber\\&&
   -\frac{1}{2 \sqrt{\omega_2^2}(\omega_2^2-\omega_1^2 )} 
   \Big(4 C D^2+\omega_2^2 D+\omega_2^2 B_2-4 C B_2 B_1\nonumber\\&&
   -4 A_2^2 B_2 (4\langle S^x_A\rangle^2+4\langle S^z_A\rangle^2-1)\nonumber\\&&
   +A_1 (4\langle S^x_B\rangle^2+4\langle S^z_B\rangle^2-1)\times\nonumber\\&&
    \big(\omega_2^2-4 C B_1-
   4 A_1^2 (4\langle S^x_A\rangle^2+4\langle S^z_A\rangle^2-1)\nonumber\\&&
   -4A_1 (B_1+C (4\langle S^x_A\rangle^2+4\langle S^z_A\rangle^2-1))\big)\nonumber\\&&
   +4 A_1 \big(D^2-B_2(B_1+C (4\langle S^x_A\rangle^2+4\langle S^z_A\rangle^2-1))\big)\Big)
\end{eqnarray}
the correlation functions for the ferromagnetic and the anti-ferromagnetic 
phases are:
\begin{eqnarray}
\lefteqn{\langle S^y_i S^y_{j\in A}(0) \rangle_k=}\nonumber\\&&
\frac{\coth (\frac{\beta  \omega_1}{2}) } {2
   (\omega_1^3-\omega_1 \omega_2^2 ) }\times\nonumber\\&&
((A_3 \langle S^z_A \rangle +B_3 \langle S^z_B \rangle)
   \omega_1^2  -(A_3 A_3-B_3
   B_4) (A_4 \langle S^z_A \rangle-B_3 \langle S^z_B \rangle))
  \nonumber\\&&
   +\frac{\coth(\frac{\beta 
   \omega_2}{2})} {2 (\omega_2^3-
   \omega_1^2 \omega_2) }\times\nonumber\\&&
((A_3
   \langle S^z_A \rangle+B_3 \langle S^z_B \rangle) \omega_2^2 
   -(A_3 A_4-B_3 B_4) (A_4
   \langle S^z_A \rangle-B_3 \langle S^z_B \rangle)) 
\end{eqnarray}
\begin{eqnarray}
\lefteqn{\langle S^y_i S^y_{j\in B}(0) \rangle_k=}\nonumber\\&&
\frac{\coth (\frac{\beta  \omega_1}{2}) } {2
   (\omega_1^3-\omega_1 \omega_2^2 ) ^4}\times\nonumber\\&&
((A_4 \langle S^z_B \rangle +B_4 \langle S^z_A \rangle)
   \omega_1^2 -(A_4 A_4-B_4
   B_3) (A_3 \langle S^z_B \rangle-B_4 \langle S^z_A \rangle))
  \nonumber\\&&
   +\frac{\coth(\frac{\beta 
   \omega_2}{2})} {2 (\omega_2^3-
   \omega_1^2 \omega_2)  }\times\nonumber\\&&
((A_4
   \langle S^z_B \rangle+B_4 \langle S^z_A \rangle) \omega_2^2 
   -(A_4 A_3-B_4 B_3) (A_3
   \langle S^z_B \rangle-B_4 \langle S^z_A \rangle)) 
\end{eqnarray}
which can be written in the zero temperature limit as:
\begin{eqnarray}
\lefteqn{\langle S^y_i S^y_{j\in A}(0) \rangle_k=}\nonumber\\&&
\frac{(A_3 A_4-B_3 B_4)(A_4 \langle S_{z_A} \rangle-B_3  \langle S_{z_B} \rangle)+
(A_3 \langle S_{z_A} \rangle-B_3  \langle S_{z_B} \rangle) \sqrt{(\omega_1 \omega_2)^2}}
{\sqrt{(\omega_1 \omega_2)^2}(\sqrt{\omega_1^2}+\sqrt{\omega_2^2})}
\end{eqnarray}
\begin{eqnarray}
\lefteqn{\langle S^y_i S^y_{j\in B}(0) \rangle_k=}\nonumber\\&&
\frac{(A_3 A_4-B_3 B_4)(A_3 \langle S_{z_B} \rangle-B_4  \langle S_{z_A} \rangle)+
(A_4 \langle S_{z_B} \rangle-B_4  \langle S_{z_A} \rangle) \sqrt{(\omega_1 \omega_2)^2}}
{\sqrt{(\omega_1 \omega_2)^2}(\sqrt{\omega_1^2}+\sqrt{\omega_2^2})}
\end{eqnarray}

\subsection{Derivation of the Matrix Equation}

Finally we carry out the derivation of the matrix equation, 
evolving from the equations of motion.
The three equations of motion for the Green's functions after employing the 
cumulant decoupling scheme become:
\begin{eqnarray}
\lefteqn{i\partial_t G^{xy}_{ij}(t)=
            i  \delta(t) \delta_{ij} \langle S^z_i(t) \rangle
           -ih_z G^{yy}_{ij}(t)}\nonumber\\
&&-2i\sum_lJ^{\|}_{il} (\langle S^y_i(t)\rangle G^{zy}_{lj}(t)+
            \langle S^z_l(t)\rangle G^{yy}_{ij}(t)) \nonumber \\
&&  +2i\sum_lJ^{\top}_{il}
           (\langle S^z_i(t)\rangle G^{yy}_{lj}(t)+
            \langle S^y_l(t)\rangle G^{zy}_{ij}(t)) 
\end{eqnarray}
\begin{eqnarray}
\lefteqn{i\partial_t G^{yy}_{ij}(t)=
ih_z G^{xy}_{ij}(t)}\nonumber\\
        &&+2i\sum_lJ^{\|}_{il}
           (\langle S^x_i(t)\rangle G^{zy}_{lj}(t)
            +\langle S^z_l(t)\rangle G^{xy}_{ij}(t)) \nonumber \\
          &&-2i\sum_lJ^{\top}_{il}
           (\langle S^z_i(t)\rangle G^{xy}_{lj}(t)+
            \langle S^x_l(t)\rangle G^{zy}_{ij}(t))
\end{eqnarray}
\begin{eqnarray}
\lefteqn{i \partial_t G^{zy}_{ij}(t)=
           -i  \delta(t) \delta_{ij} \langle S^x_i(t) \rangle }\nonumber\\
           &&+2i\sum_lJ^{\top}_{il}
           (\langle S^y_i(t)\rangle G^{xy}_{lj}(t)+
            \langle S^x_l(t)\rangle G^{yy}_{ij}(t)) \nonumber \\
           &&-2i\sum_lJ^{\top}_{il}
           (\langle S^x_i(t)\rangle G^{yy}_{lj}(t)+
            \langle S^y_l(t)\rangle G^{xy}_{ij}(t))           
\end{eqnarray}
Fourier transforming into $\omega$-space and assume that the spins
are constants in time:
\begin{eqnarray}
\lefteqn{ \omega G^{xy}_{ij}(\omega)=
            \frac{i}{2\pi}\delta_{ij} \langle S^z_i \rangle
           -ih_z G^{yy}_{ij}(\omega)}\nonumber\\
           &&-2i\sum_lJ^{\|}_{il}
           (\langle S^y_i\rangle G^{zy}_{lj}(\omega)+
            \langle S^z_l\rangle G^{yy}_{ij}(\omega)) \nonumber \\
           &&
           +2i\sum_lJ^{\top}_{il}
           (\langle S^z_i\rangle G^{yy}_{lj}(\omega)+
            \langle S^y_l\rangle G^{zy}_{ij}(\omega)) 
\end{eqnarray}
\begin{eqnarray}
\lefteqn{ \omega G^{yy}_{ij}(\omega)=
           ih_z G^{xy}_{ij}(\omega)}\nonumber\\
       && +2i\sum_lJ^{\|}_{il}
           (\langle S^x_i\rangle G^{zy}_{lj}(\omega)+
            \langle S^z_l\rangle G^{xy}_{ij}(\omega)) \nonumber \\
           &&-2i\sum_lJ^{\top}_{il}
           (\langle S^z_i\rangle G^{xy}_{lj}(\omega)+
            \langle S^x_l\rangle G^{zy}_{ij}(\omega)) 
\end{eqnarray}
\begin{eqnarray}
\lefteqn{ \omega G^{zy}_{ij}(\omega)=
           -\frac{i}{2\pi} \delta_{ij} \langle S^x_i \rangle}\nonumber\\
          && +2i\sum_lJ^{\top}_{il}
           (\langle S^y_i\rangle G^{xy}_{lj}(\omega)+
            \langle S^x_l\rangle G^{yy}_{ij}(\omega)) \nonumber \\
           &&
           -2i\sum_lJ^{\top}_{il}
           (\langle S^x_i\rangle G^{yy}_{lj}(\omega)+
            \langle S^y_l\rangle G^{xy}_{ij}(\omega))           
\end{eqnarray}
In the canted anti-ferromagnetic and the anti-ferromagnetic 
phases the spins on the two sub-lattices assume different values.
Therefore we need to split up every Green's function before 
we can Fourier Transform into k-space.
We define $G_A^{\mu\nu}(k,\omega)$ and $G_B^{\mu\nu}(k,\omega)$.
$G_A^{\mu\nu}(k,\omega)$  refers to $G_{ij}^{\mu\nu}(\omega)$  when
the site $i$ is on sub-lattice A
and $G_B^{\mu\nu}(k,\omega)$ when $i$ is on sub-lattice B.
\begin{eqnarray}
\lefteqn{ \omega {G_A}^{xy}_{k}(\omega)=
            \frac{i}{(2\pi)^4}  \langle S^z_A \rangle
            -2ih_z {G_A}^{yy}_{k}(\omega)}\nonumber\\
            &&- 4i \langle S^y_A \rangle
             [{G_A}^{zy}_{k}(\omega)J^{\|}_2 \gamma_2 (k)
            +{G_B}^{zy}_{k}(\omega)J^{\|}_1 \gamma_1 (k)]
             \nonumber \\
           &&-4i {G_A}^{yy}_{k}(\omega) 
             [\langle S^z_A \rangle J^{\|}_2 \gamma_2 (0)
           +\langle S^z_B \rangle J^{\|}_1  \gamma_1 (0) ]  \nonumber \\
           &&+4i \langle S^z_A \rangle
            [{G_A}^{yy}_{k}(\omega)J^{\top}_2 \gamma_2 (k)
           +{G_B}^{yy}_{k}(\omega)J^{\top}_1 \gamma_1 (k)  ]  \nonumber \\
           &&+4i {G_A}^{zy}_{k}(\omega) 
             [\langle S^y_A \rangle  J^{\top}_2 \gamma_2 (0)
             +\langle S^y_B \rangle J^{\top}_1 \gamma_1 (0) ] \nonumber \\
\end{eqnarray}
\begin{eqnarray}
\lefteqn{     \omega {G_A}^{yy}_{k}(\omega)=
             -2ih_z {G_A}^{xy}_{k}(\omega)}\nonumber\\&&+       
           4i \langle S^x_A \rangle
             [{G_A}^{zy}_{k}(\omega) J^{\|}_2 \gamma_2 (k)
           +{G_B}^{zy}_{k}(\omega) J^{\|}_1 \gamma_1 (k) ]  \nonumber \\
          && +4i {G_A}^{xy}_{k}(\omega) 
             [\langle S^z_A \rangle J^{\|}_2 \gamma_2 (0)
           +\langle S^z_B \rangle J^{\|}_1  \gamma_1 (0) ]  \nonumber \\
           &&-4i \langle S^z_A \rangle
            [{G_A}^{xy}_{k}(\omega)J^{\top}_2 \gamma_2 (k)
           +{G_B}^{xy}_{k}(\omega)J^{\top}_1 \gamma_1 (k)  ]  \nonumber \\
           &&-4i {G_A}^{zy}_{k}(\omega) 
             [\langle S^x_A \rangle  J^{\top}_2 \gamma_2 (0)
             +\langle S^x_B \rangle  J^{\top}_1  \gamma_1 (0)]  \nonumber \\
\end{eqnarray}
\begin{eqnarray}
\lefteqn{ \omega {G_A}^{zy}_{k}(\omega)=
            -\frac{i}{(2\pi)^4}  \langle S^x_A \rangle} \nonumber\\
            &&+4i \langle S^y_A \rangle
             [{G_A}^{xy}_{k}(\omega) J^{\top}_2 \gamma_2 (k)
           +{G_B}^{xy}_{k}(\omega)  J^{\top}_1 \gamma_1 (k)] \nonumber \\
           &&-4i {G_A}^{yy}_{k}(\omega) 
             [\langle S^x_A \rangle  J^{\top}_2 \gamma_2 (0)
           +\langle S^x_B \rangle J^{\top}_1  \gamma_1 (0)]\nonumber \\
           &&-4i \langle S^x_A \rangle
            [{G_A}^{yy}_{k}(\omega)J^{\top}_2 \gamma_2 (k)
           +{G_B}^{yy}_{k}(\omega)J^{\top}_1 \gamma_1 (k) ]  \nonumber \\
           &&-4i {G_A}^{xy}_{k}(\omega) 
             [\langle S^y_A \rangle  J^{\top}_2 \gamma_2 (0)
             +\langle S^y_B \rangle  J^{\top}_1  \gamma_1 (0)] 
\end{eqnarray}

\section{Conclusions}

In conclusion, we analysed the anisotropic Heisenberg model in a external field
on the three dimensional bcc lattice by employing 
the well-established technique of real-time Green's functions
for spin systems. The series of infinite order Green's functions as it appears 
in the equation of motion was truncated by applying cumulant decoupling and
the resulting random phase approximation accounts for linear spin-waves. 
We are the first to apply this method to the canted anti-ferromagnetic 
phase entailing a set of six algebraic equations.  The innate self-consistency 
equations herein constitute a three dimensional numerical integral over the k-space. 
By introducing a two dimensional density of states  the integral was reduced to 
two dimensions where the lattice generating functions serve as new
integration variables. In the appearing integrals the DOS is the only
quantity that depends on the structure of the lattice. Hence, once
the DOS is computed for a certain lattice geometry the further
calculation remain unaltered. Therefore our method is widely applicable
and easily adjustable to various magnetic systems  where 
canted phases are in the center of interest. This also holds for
two dimensional lattices where linear spin waves are expected to yield a
reasonable approximation. 

\section{Acknowledgements}

One of the authors (Mikl\'{o}s Gul\'{a}csi) wishes to thank James L. Smith 
for his friendship, guidance and encouragement over the years. Thanks for 
the "{\it{shmokos}}"!

\appendix

\section{Connection to ${}^4$He physics}

For the interested reader, we briefly summarize the connection between the 
anisotropic Heisenberg model and ${}^4$He as introduced by Matsubara and Matsuda  
\cite{MatsubaraMatsuda1,MatsubaraMatsuda2}, and used by Matsuda and Tsuneto
\cite{MatsudaTsuneto}, Fisher \cite{Fisher}, Liu and Fisher
\cite{LiuFisher} and most recently by Stoffel and Gul\'{a}csi 
\cite{epl,epjb1,epjb2}. 

Apart from possible ${}^3$He impurities ${}^4$He is  
a bosonic system and the generic Hamiltonian for such systems 
in the language of second quantization is given by:
\begin{eqnarray} \label{genbosH}
H&=&\int d^3 x \hat{\psi}^{\dagger}({\bf{x}})( -\frac{1}{2 m }\nabla^2+
\mu) \hat{\psi}({\bf{x}})  \nonumber\\
 &&+\frac{1}{2}\int d^3x d^3x'\hat{\psi}^{\dagger}({\bf{x}})
 \hat{\psi}^{\dagger}({\bf{x}}') V({\bf{x}}-{\bf{x}}')
 \hat{\psi}({\bf{x}})\hat{\psi}({\bf{x}}') \nonumber \\
\end{eqnarray}
where $\psi^{\dagger}({\bf{x}})$, the particle creation operator and 
$\psi^{\dagger}({\bf{x}})$ , the corresponding destruction operator 
obey the usual bosonic commutator relations.
Hamiltonians in three dimensions such as in Equation (\ref{genbosH}) are not solvable 
even for elementary potentials $V({\bf{x}})$ such as the Dirac delta distribution.
Therefore, further approximations has to be implemented. 
An approximation which proved particularly successful 
for the description of liquid Helium is know
as the {\sl{ quantum lattice gas}} model and was first introduced
by Matsubara and Matsuda \cite{MatsubaraMatsuda1,MatsubaraMatsuda2}.

In the quantum lattice gas model one works with a space lattice of discrete 
lattice points rather than the continuum.  This approximation proves 
to be very useful to study solid states as the spacial discretization of 
this model serves as a natural frame for the crystal lattice. Also in this 
model no specific knowledge of the density distribution of the atoms is needed.

According to Matsubara and Tsuneto \cite{MatsudaTsuneto} the generic 
Hamiltonian Equation (\ref{genbosH}) in the discrete lattice model reads:
\begin{eqnarray}\label{hcbhubb}
H=\mu \sum_i n_i+
\sum_{ij}u_{ij}(a_i^{\dagger}-a_j^{\dagger})
(a_i-a_j)
+\sum_{ij} V_{ij} n_i n_j \nonumber \\
\end{eqnarray}
Here $u_{ij}$ are non-zero for nearest neighbor and next nearest neighbor 
hopping and otherwise zero. The values of $u_{nn}$ and $u_{nnn}$ are  such
that  the kinetic energy is isotropic up to the 4th order. In the case of 
a bcc lattice (two interpenetrating sc lattices) the matrix elements are 
given by:
\begin{eqnarray}
u_{nn}=\frac{2}{3}\frac{}{4 m a^2}\\
u_{nnn}=\frac{1}{3}\frac{}{4 m  a^2}
\end{eqnarray}
As the atoms do not penetrate each other there can exist only one atom at a 
time on a lattice site. Consequently $a^{\dagger}$ and $a$ are the creation 
and annihilation operators of a hard core boson commuting on different lattice 
sites:
\begin{eqnarray}
[a^{\dagger}_i,a^{\dagger}_j]_{-}=
[a_i,a_j]_{-}=
[a_i,a^{\dagger}_j]_{-}=0 \; (i\neq j)
\end{eqnarray}
but obey the anti-commutator relations on identical sites:
\begin{eqnarray}
[a^{\dagger}_i,a^{\dagger}_i]_{+}=
[a_i,a_i]_{+}=0   \nonumber \\
 \left[a_i,a^{\dagger}_i\right]_{+}=1 
\end{eqnarray}
Equation (\ref{hcbhubb}) is the Bose-Hubbard model in three dimensions for hard 
core bosons. Due to the unusual statistics of hard core bosons, Wick theorem 
cannot be applied  and hence, the common formalism of perturbative field 
theory is not applicable. The way out is to transform the model to an 
equivalent spin model \cite{MatsubaraMatsuda1,MatsubaraMatsuda2}, namely 
by using 
\begin{eqnarray}
a^{\dagger}_j=S^x_j-i S^y_j \nonumber  \\
a_j=S^x_j+i S^y_j    \nonumber \\
n_j=\frac{1}{2}- S^z_j
\end{eqnarray}
It can be verified that the usual Lie algebra for spin 1/2 particles preserves 
the mixed commutation/anti-commutation relations for hard-core bosons. This 
substitution transforms the hard-core bosonic Hubbard model into a spin model:
\begin{eqnarray}
\lefteqn{H=\mu\sum_i (\frac{1}{2}-S^z_i)}\nonumber\\
&&+\sum_{ij}u_{ij}
(1-S^z_i-S^z_j-2S^x_i S^x_j-2S^y_iS^y_j) \nonumber \\
&&+\sum_{ij}V_{ij}(\frac{1}{4}-\frac{S^z_i}{2}-\frac{S^z_j}{2}+S^z_iS^z_j)
\end{eqnarray}
If we adjust the notation to conform with the usual standards of spin models, we 
re-obtain the anisotropic Heisenberg from Equation (\ref{new_one}): 
\begin{eqnarray}
H=h^z \sum_i S^z_i+\sum_{ij}J^{\|}_{ij}S^z_iS^z_j+
\sum_{ij}J^{\top}_{ij}(S^x_i S^x_j+S^y_iS^y_j) 
\end{eqnarray}
with:
\begin{eqnarray}
J^{\|}_{ij}=V_{ij}\nonumber \\
J^{\top}_{ij}=-2 u_{ij}
\nonumber\\
h^z=-\mu+\sum_{j}J^{\top}_{ij}
-\sum_{j}J^{\|}_{ij}
\end{eqnarray}

If the above presented transformation is used for ${}^4$He, then 
the values of the $J$'s also have to be chosen such as to mimic ${}^4$He. 
The interactions between the ${}^4$He atoms are controlled
by van-der-Waals forces and their repulsive nature at very short distances
determines negative nearest neighbor interaction $J^{\|}_1$, evoking
anti-ferromagnetic ordering in the spin language. The corresponding 
Lennard-Jones potential is short ranged and therefore it is sufficient \cite{LiuFisher}
to only consider nearest and next nearest neighbor interactions. Hence, for
${}^4$He the $J$ values will be 
$J^{\|}_1=-q_1 J^{\|}_{i\in A j\in B} $, $J^{\|}_2=-q_2 J^{\|}_{i\in A j\in A}$, 
$J^{\top}_1=-q_1 J^{\top}_{i\in A j\in B}$ and 
$J^{\top}_2=-q_2 J^{\top}_{i\in A j\in A}$, 
where $q_1=6$ and $q_2=8$ are the number of nearest and next nearest neighbors 
on the bipartite bcc lattice. Liu and Fisher \cite{LiuFisher} in their calculations 
used, $J^{\top}_1 = 1.4K$, $J^{\top}_2 = 0.5K$, $J^{\|}_1 = -3.8K$ and 
$J^{\|}_2 = -1.7K$. In a ${}^4$He calculation the results do not change 
\cite{epjb1,epjb2} if the $J^{\|}$'s are within $\pm 2$ range of these values 
and $J^{\|}_{B} > J^{\|}_{A}$ and $J^{\top}$'s values remain positive. 

Defining two sub-lattices gave 
\cite{Fisher,LiuFisher,MatsubaraMatsuda1,MatsubaraMatsuda2,MatsudaTsuneto}
a possibility to establish the diagonal long-range order
of solids in a natural way: sub-lattice A represents the centers of the 
${}^4$He ions, hence it coincides with the ion lattice. Sub-lattice B
defines the interstitials, the space in-between those atomic centers.
In the liquid phases, of course, the occupation number on both 
sub-lattices is equal as there is no spacial density variation.
In Table (\ref{tab:one}) we gave the various magnetic phases 
of the anisotropic Heisenberg model. These phases, however 
identify the corresponding phases of the ${}^4$He system,
as presented in Table (\ref{tab:two}). 

\begin{table}
\centering
\begin{tabular}{|c|c|c|c|c|}
\hline
&&&&\\
Spin Configuration& Magnetic Phase & ~~~ ODLRO ~~~  & ~~~~ DLRO ~~~~ & ${}^4$He-Phase \\
&&&&\\
\hline
&&&&\\
$\uparrow\uparrow$ & FE & No & No & Normal Liquid \\
&&&&\\
$\nearrow\nearrow$ & CFE & Yes & No & Superfluid \\
&&&&\\
$\nearrow\swarrow$ & CAF  & Yes & Yes & Supersolid \\
&&&&\\
$\uparrow\downarrow$& AF & No & Yes & Normal Solid \\
&&&&\\
\hline
\end{tabular}
\caption{\label{tab:two} The phases of ${}^4He$ corresponding to
the phases of the anisotropic Heisenberg model. Similarly to Table (\ref{tab:one}) 
the phases are defined by their long range order, i.e., 
off-diagonal long-range order (ODLRO) and diagonal long-range order (DLRO).}
\end{table}

\section{The mean-field limit}

In the mean-field solution of Liu and Fisher \cite{LiuFisher} the 
Green's functions have not been explicitly evaluated, for the sake 
of completeness we re-derive these hereafter. 

First we will re-derive the classical mean-field approximation as 
was pioneered by Liu and Fisher \cite{LiuFisher} and briefly state 
some key properties. We will further show that this approximation 
is a special case of the random-phase approximation. 

The anisotropic Heisenberg Hamiltonian in the classical
mean-field approximation is obtained by substituting the 
spin $1 / 2$ operators with their respective 
expectation values:
\begin{eqnarray}
 \lefteqn{ H_{MF}=- h^z  (\langle S^z_A \rangle +
    \langle S^z_B \rangle)}\nonumber\\
&&-2 J^{\|}_1 \langle S^z_A\rangle\langle S^z_B\rangle
-J^{\|}_{2}(\langle S^z_A\rangle \langle S^z_A \rangle + 
\langle S^z_B\rangle \langle S^z_B\rangle) \nonumber\\
&&-2 J^{\top}_1 \langle S^x_A\rangle\langle S^x_B\rangle
-J^{\top}_{2}(\langle S^x_A\rangle \langle S^x_A \rangle 
+ \langle S^x_B\rangle \langle S^x_B\rangle ) \nonumber\\
\label{HMF}
\end{eqnarray}
Here $J^{\|}_1=-q_1 J^{\|}_{i\in A,j\in B} $, $J^{\|}_2=-q_2 J^{\|}_{i\in A,j\in A}$, 
$J^{\top}_1=-q_1 J^{\top}_{i\in A,j\in B}$ and $J^{\top}_2=-q_1 J^{\top}_{i\in A,j\in A}$
where $q_1=6$ and $q_2=8$ are the number of nearest and 
next nearest neighbours on the bipartite bcc lattice.
The mean value of $S_y$ drops out as the randomly broken symmetry
$S_x \leftrightarrow S_y$ (off-diagonal long-range order) 
allows for $\langle S_y \rangle=0$. 
The standard method of deriving the corresponding self-consistency 
equations is to minimize the Helmholtz's Free energy $F=H-TS$. 
The entropy $S$ is given by the pseudo spin entropy of the system:
\begin{eqnarray}\label{pspin}
S=-\frac{1}{2}[(\frac{1}{2}+S_A)\ln(\frac{1}{2}1+S_A)+(\frac{1}{2}-S_A)\ln(\frac{1}{2}-S_A) \nonumber \\
+(\frac{1}{2}+S_B)\ln(\frac{1}{2}+S_B)+(\frac{1}{2}-S_B)\ln(\frac{1}{2}-S_B)]
\end{eqnarray}
where $S_A=\sqrt{\langle S_{z_A} \rangle^2+\langle S_{x_A}\rangle^2}$ and
 $S_B=\sqrt{\langle S_{z_B}\rangle^2 +\langle S_{x_B}\rangle^2 }$.
We could equally well say that the state of the system is determined by 
minimizing the internal energy $\langle H \rangle$, subject to an additional 
constraint given by Equation (\ref{pspin}). In this picture the temperature 
becomes a Lagrange multiplier and 
at absolute zero, where $T=S=0$ we obtain, as $\lim_{S_A,S_B\rightarrow 0} S=0$,
\begin{equation}
\sqrt{\langle  S^x_A\rangle^2 +\langle S^z_A \rangle^2 }=
\sqrt{\langle  S^x_B\rangle^2 +\langle S^z_B \rangle^2 }=\frac{1}{2}
\end{equation}
This result implies that this approximation does not take quantum fluctuations 
into account.

In the canted anti-ferromagnetic and the canted ferromagnetic states there are 
four self-consistency equations
in the ferromagnetic and anti-ferromagnetic phases; where 
$\langle S_x\rangle=\langle S_y\rangle=0$ they are
reduced in number by two. These equations are readily obtained by differentiating
 the free energy with respect to $\langle S_z\rangle$ and
 $\langle S_x\rangle$ respectively.
The resulting equations can be rearranged to yield:
\begin{eqnarray}{\label{mf1}}
\langle S^x_A\rangle= \frac{2 J^{\top}_1\langle S^x_B\rangle
          +2 J^{\top}_2 \langle S^x_A \rangle}{2\omega_A} \tanh(\beta\omega_A)\nonumber \\
\langle S^z_A\rangle= \frac{2 J^{\|}_1\langle S^z_B\rangle
          +2 J^{\|}_2 \langle S^z_A \rangle+h^z}{2\omega_A} \tanh(\beta\omega_A) \nonumber \\
\langle S^x_B\rangle= \frac{2 J^{\top}_1\langle S^x_A\rangle
          +2 J^{\top}_2 \langle S^x_B \rangle}{2\omega_B} \tanh(\beta\omega_B) \nonumber \\
\langle S^z_B\rangle= \frac{2 J^{\|}_1\langle S^z_A\rangle
          +2 J^{\|}_2 \langle S^z_B \rangle+h^z}{2\omega_B} \tanh(\beta\omega_B) 
\end{eqnarray}
where
\begin{eqnarray}\label{cmfw}
\omega_A=[(2 J^{\top}_1\langle S^x_B\rangle
          +2 J^{\top}_2 \langle S^x_A \rangle)^2+\nonumber\\
      (2 J^{\|}_1\langle S^z_B\rangle
          +2 J^{\|}_2 \langle S^z_A \rangle+h^z)^2]^{\frac{1}{2}} \nonumber\\
\omega_B=[(2 J^{\top}_1\langle S^x_A\rangle
          +2 J^{\top}_2 \langle S^x_B \rangle)^2+\nonumber\\
      (2 J^{\|}\langle S^z_B\rangle
          +2 J^{\|} \langle S^z_A \rangle+h^z)^2]^{\frac{1}{2}}
\end{eqnarray}
In the canted phases where the transversal magnetic fields 
$\langle S^x_A \rangle $ and $\langle S^x_B\rangle $ are non-zero, the energies 
$\omega_A$ and $\omega_B$ can be eliminated from 
 equations (\ref{mf1}) to yield the following important 
relations:
\begin{eqnarray} \label{mf2}
   h^z+2 \langle S^z_A \rangle (J_2^{\|}-J_2^{\top})+2 \langle S^z_B \rangle J_1^{\|}=
   2 J_1^{\top}\frac{\langle S^x_B\rangle }{\langle S^x_A\rangle }\langle S^z_A\rangle 
   \nonumber \\
   h^z+2 \langle S^z_B\rangle  (J_2^{\|}-J_2^{\top})+2 \langle S^z_A \rangle J_1^{\|}=
   2 J_1^{\top}\frac{\langle S^x_A\rangle }{\langle S^x_B\rangle }\langle S^z_B \rangle \nonumber\\
\end{eqnarray}

In the limit $h^z \rightarrow \infty$ 
 the Hamiltonian (Equation (\ref{HMF})) reduces to an 
effective single operator model:
\begin{equation}\label{fs}
H=-h^z (\langle S^z_A\rangle +\langle S^z_B\rangle )
\end{equation}
Consequently for high external fields the system 
will assume the energetically favorable ferromagnetic phase.
In the opposite limit  $h^z \rightarrow 0$ and 
with sufficiently large anti-ferromagnetic 
nearest neighbour coupling 
$J^{\|}_1\ll 0$ the system is dominated by:  
\begin{equation}\label{afs}
H=-J^{\|}_1 \langle S^z_A\rangle\langle S^z_B\rangle
\end{equation}
giving rise to the anti-ferromagnetic state. 
At medium large fields $h^z$ the two terms
Equation (\ref{fs}) and Equation (\ref{afs}) 
balance each other and the transversal 
ferromagnetic coupling  ($J^{\top}_1>0$ 
and $J^{\top}_2>0$) become significant, 
deviating the spins into the $x$-direction.
In regions of higher $h^z$, where 
ferromagnetism (Equation (\ref{fs})) is more prevalent the system 
leaps into the canted ferromagnetic phase;
for lower $h^z$ it yields the canted anti-ferromagnetic phase.
For those sets of coupling constant where
all four phases are existent, the 
corresponding phase transitions are of second order. 
If, due to choice of constants one or more of 
those phases, for example
the canted anti-ferromagnetic phase, does not appear
 the resulting
canted ferromagnetic to ferromagnetic phase 
transition is first order.

Now, we may ask the question "For which sets 
of parameters all  four phases appear?". 
Matsuda and Tsuneto \cite{MatsudaTsuneto} derived  relations for all 
phase transition points at zero temperature.
As mentioned in the previously
the four phases are distinguished by their
order parameters, $m_1=\langle S^x_A\rangle+\langle S^x_B\rangle $ and 
$m_2=\langle S^z_A\rangle-\langle S^z_B\rangle$. 
Across all second order phase transitions the 
spin mean-fields $\langle S_x \rangle $ and $\langle S_z \rangle $ are continuous
$C^0$ functions of $h^z$. Therefore the critical 
points are determined by Equations (\ref{mf2}), 
in the limits where the relevant order parameter,
$m_1$ or $m_2 $, disappears.
The canted ferromagnetic phase transits into 
the ferromagnetic phase when $m_1$ approaches zero.
Hence we set $\langle S^x_A\rangle =\langle S^x_B\rangle \rightarrow 0$ and consequently
 $\langle S^z_A\rangle =\langle S^z_B\rangle =\frac{1}{2}$.
Equation (\ref{mf2}) readily gives:
\begin{equation}\label{tpfsf}
h^z_{FE-CFE}=J^{\top}_1+J^{\top}_2-J^{\|}_1-J^{\|}_2
\end{equation}
Equally the 
canted anti-ferromagnetic to anti-ferromagnetic transition
is defined by $\langle S^z_A\rangle =-\langle S^z_B\rangle =\frac{1}{2}$ 
while $\langle S^x_A\rangle\rightarrow 0$ and $
 \langle S^x_B\rangle \rightarrow 0$. The unknown quotient 
$\frac{\langle S^x_A\rangle }{\langle S^x_B\rangle }$ is readily eliminated:
\begin{equation}\label{tpiaf}
h^z_{CAF-AF}=\sqrt{(-J^{\|}_1+J^{\|}_2-J^{\top}_2)^2-(J^{\top}_1)^2}
\end{equation}
The canted ferromagnetic and the canted anti-ferromagnetic phases 
coexist where the order parameter of the diagonal long-range order, 
$m_2=\langle S^z_A\rangle-\langle S^z_B\rangle $ approaches zero.  
We replace $\langle S^z_A\rangle $ 
and $\langle S^z_B \rangle$ in equation (\ref{mf2}) with $m_2$ and 
$m_1$
and retain only linear terms of $m_2$. Subtracting and summing up both 
equations respectively yields:
\begin{eqnarray}
   h^z+2 m_2(J_2^{\|}-J_2^{\top}+J_1^{\|})=
     2 J_1^{\top}m_2
   \nonumber \\
   2 m_1  (J_2^{\|}-J_2^{\top}- J_1^{\|})=
   -2 J_1^{\top} m1 \frac{ 4m_2^2+1}{4 m_2^2-1}
\end{eqnarray}
We used that $\sqrt{\langle S^z_A\rangle^2+\langle S^x_A\rangle^2}=\frac{1}{4}$ at T=0.
The solution of these two equations determine
the critical point which is given by:
\begin{eqnarray}\label{tpisf}
h^z_{CFE-CAF}=\frac{J^{\|}_1+J^{\|}_2-J^{\top}_1-J^{\top}_2}
{J^{\|}_1-J^{\|}_2-J^{\top}_1+J^{\top}_2}\times\nonumber \\
\sqrt{(-J^{\|}_1+J^{\|}_2-J^{\top}_2)^2-(J^{\top}_1)^2}
\end{eqnarray}
For a particular choice of coupling constants all four phases 
will exists when:
\begin{eqnarray}
h^z_{FE-CFE}>h^z_{CFE-CAF}>h^z_{CAF-AF}
\end{eqnarray}
In other cases, for example where 
$h^z_{CFE-CAF}<h^z_{CAF-AF}$ the canted anti-ferromagnetic
phase is suppressed. The resulting first order canted ferromagnetic -
anti-ferromagnetic transition point has to be calculated by 
finding the state with the lowest 
internal energy $\langle H \rangle$ and making a comparison between the two.

\section{Link to Mean-Field Solution}

Here we establish a link between the classical
mean-field approximation and the random-phase approximation as derived in the 
previous Chapters.
We have already seen that both approximations
are mean-field type, involving self consistency equations and that there are two 
equations (Equation (\ref{mf2}) or Equation (\ref{sssfcon})) which appear
in both approximations. However, the random-phase approximation
takes spin-wave/quasi-particle excitations into account whereas the 
classical mean-field approximation is an effective one operator
model exhibiting two energy levels per sub-lattice.
Therefore we can now consider if those energy levels, specifically their difference,
correspond to certain spin-wave excitations.
In this Appendix will show that the classical mean-field approximation
is a limiting case of the random-phase approximation in a way in which the 
integral of the momentum $k$ is restricted to values where:
\begin{eqnarray}
\gamma_1(k)=
\gamma_2(k)=0, 
\end{eqnarray}
or equally the generalized density of states is:
\begin{eqnarray}
 \rho(\gamma_1,\gamma_2)=\delta(\gamma_1)\delta(\gamma_2)
\end{eqnarray} 
In this limit the self-consistency equation of the random-phase approximation 
Equation (\ref{eq:selfconftfg}) becomes
\begin{eqnarray}\label{eq:selfconfmf}
F_{A_{MF}}&=&\sum_{\omega_p}[\mbox{Residue}
  (\frac{G^{yy}_{A}(\gamma_1=0,\gamma_2=0,\omega)}{e^{\beta \omega}-1},-\omega_p)\nonumber \\
 && +\mbox{Residue}(\frac{G^{yy}_{A}(\gamma_1=0,\gamma_2=0,\omega)}{e^{\beta \omega}-1},-\omega_p)]
        -\frac{1}{4}=0 \nonumber\\
F_{B_{MF}}&=&\sum_{\omega_p}[\mbox{Residue}
 (\frac{G^{yy}_{B}(\gamma_1=0,\gamma_2=0,\omega)}{e^{\beta \omega}-1},-\omega_p)\nonumber \\
  && +\mbox{Residue}(\frac{G^{yy}_{B}(\gamma_1=0,\gamma_2=0,\omega)}{e^{\beta \omega}-1},-\omega_p)]
  -\frac{1}{4}=0  
\end{eqnarray}
\begin{figure}[t]
\centering
\includegraphics[width=10cm]{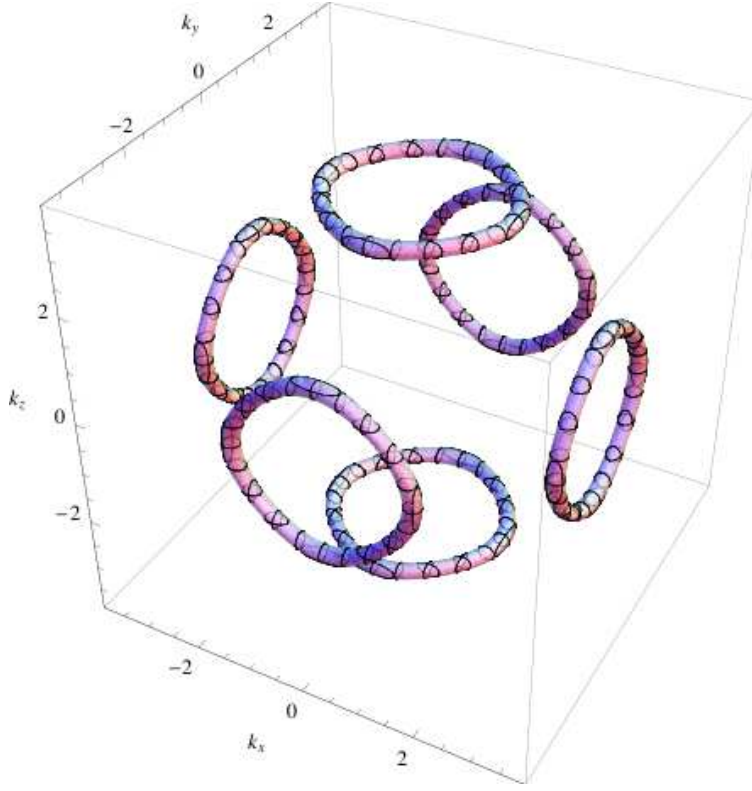} 
\caption{\label{fig:four} The rings show the areas where the two
lattice generating functions $\gamma_1(k)$ and $\gamma_2(k)$
yield simultaneously zero. If the integral over the first Brillouin zone
is restricted to those areas the random-phase approximation devolves into 
the classical mean-field approximation.}
\end{figure} 
Figure \ref{fig:four} shows the wave-vectors $k$ that correspond to $\gamma_1(k)=
\gamma_2(k)=0$ within the first Brillouin zone.
Physically, vanishing lattice generating functions ($\gamma_1(k)=\gamma_2(k)=0$)
 means that the system loses
all information about the lattice structure as it is the case in the 
classical mean-field approximation where the only information that remains
is the number of nearest and next nearest neighbors.
With the lattice generation functions being zero the matrix $M$ of Equation 
(\ref{Meq}) becomes:
\begin{eqnarray}
M^{\tiny{\mbox{MF}}}=\left(
\begin{array}{rrrrrr}
 i\omega & 0    &M_{13}&0&0&0 \\
 0     & i\omega&0&M_{24}&0&0 \\ 
-M_{13}&0&i\omega &0&M_{35}&0 \\
0&-M_{24}&0&i\omega &0&M_{46} \\
 0&0&M_{53}&0&i\omega &0 \\
 0&0&0&M_{64}&0&i\omega \\
\end{array}
\right)
\end{eqnarray}
where the components are given by:
\begin{eqnarray}
&&M_{13}= 2h^z+4\langle S^z_A\rangle J_2^{\|}+4\langle 
        S^z_B\rangle J_1^{\|}\nonumber\\ 
&&M_{24}=2h^z+4\langle S^z_B\rangle J_2^{\|}+4\langle 
        S^z_A\rangle J_1^{\|}\nonumber\\
&&M_{35}=4\langle S^x_A\rangle J_2^{\top}+4\langle
      S^x_B\rangle J_1^{\top}\nonumber \\
&&M_{46}=-4\langle S^x_B\rangle J_2^{\top} 
     +4\langle S^x_A\rangle J_1^{\top}\nonumber\\
&&M_{53}= -4\langle S^x_B\rangle J_1^{\top}-4\langle S^x_A\rangle
   J_2^{\top} \nonumber\\
&&M_{64}= -4\langle S^x_A\rangle 
 J_1^{\top}-4\langle S^x_B\rangle  J_2^{\top}\nonumber\\
\end{eqnarray}
The corresponding Green's functions are given by:
\begin{eqnarray}
G^{yy}_{A_{MF}}(\omega)=\frac{(4 J^{\top}_1\langle S^x_B\rangle
          +4 J^{\top}_2 \langle S^x_A \rangle) 
  \langle S^x_A \rangle+
      (4 J^{\|}_1\langle S^z_B\rangle
          +4 J^{\|}_2 \langle S^z_A \rangle+2h^z) \langle S^z_A\rangle}{\omega^2-(2\omega_1)^2}\nonumber \\
G^{yy}_{B_{MF}}(\omega)=\frac{(4 J^{\top}_1\langle S^x_A\rangle
          +4 J^{\top}_2 \langle S_{xB} \rangle) 
  \langle S^x_B \rangle+
      (4 J^{\|}_1\langle S^z_A\rangle
          +4 J^{\|}_2 \langle S^z_B \rangle+2 h^z) \langle S^z_B\rangle}{\omega^2-(2\omega_2)^2}
\end{eqnarray}
The poles of the Green's functions are given by the eigenvalues of
$M^{\mathbf{cmf}}$:
\begin{eqnarray}
\omega_1=\pm [(2 J^{\top}_1\langle S^x_B\rangle
          +2 J^{\top}_2 \langle S^x_A \rangle}{\omega_A)^2+\nonumber\\
      (2 J^{\|}_1\langle S^z_B\rangle
          +2 J^{\|}_2 \langle S^z_A \rangle+h^z)^2]^{\frac{1}{2}} \nonumber\\
\omega_2=\pm [(2 J^{\top}_1\langle S^x_A\rangle
          +2 J^{\top}_2 \langle S^x_B \rangle}{\omega_A)^2+\nonumber\\
      (2 J^{\|}\langle S^z_B\rangle
          +2 J^{\|} \langle S^z_A \rangle+h^z)^2]^{\frac{1}{2}}
\end{eqnarray}
Those energies are, as we have expected, identical to the 
classical mean field energies given by Equation (\ref{cmfw}).
The self-consistency equations calculated with Equation (\ref{eq:selfconfmf}) yield:
\begin{eqnarray}
\frac{1}{2\omega_1} [(2 J^{\top}_1\langle S^x_B\rangle
          +2 J^{\top}_2 \langle S^x_A \rangle) 
  \langle S^x_A \rangle+\nonumber\\
      (2 J^{\|}_1\langle S^z_B\rangle
          +2 J^{\|}_2 \langle S^z_A \rangle+h^z) \langle S^z_A\rangle]=
\frac{\tanh(\beta\omega)}{4}\\
  \frac{1}{2\omega_2} [(2 J^{\top}_1\langle S^x_A\rangle
          +2 J^{\top}_2 \langle S_{xB} \rangle) 
  \langle S^x_B \rangle+\nonumber\\
      (2 J^{\|}_1\langle S^z_A\rangle
          +2 J^{\|}_2 \langle S^z_B \rangle+h^z) \langle S^z_B\rangle]=
\frac{\tanh(\beta\omega)}{4}
\end{eqnarray}
Those two equations together with Equation (\ref{sssfcon}) are
identically to Equation (\ref{mf1}). Therefore we have shown that 
the mean-field as derived by Fisher and Liu \cite{LiuFisher}
is indeed a limiting case of the random-phase approximation.

\end{document}